\documentclass[a4paper,USenglish,autoref]{lipics-v2021}
\hideLIPIcs
\pdfoutput=1
\nolinenumbers

\EventEditors{Sayaka Kamei and Pierre Sutra}
\EventNoEds{2}
\EventLongTitle{30th International Conference on Principles of Distributed Systems (OPODIS 2026)}
\EventShortTitle{OPODIS 2026}
\EventAcronym{OPODIS}
\EventYear{2026}
\EventDate{December 9--11, 2026}
\EventLocation{Madrid, Spain}
\EventLogo{}
\SeriesVolume{401}
\ArticleNo{12}

\usepackage[noresetcount,vlined]{pwalgorithm}
\usepackage{pwmath}

\usepackage[compress]{cite} %

\newcommand{\defop}[1]{%
    \expandafter\newcommand\csname #1\endcsname[1]{\text{\FuncSty{#1(\ensuremath{##1})}}}%
}

\defop{Read}
\defop{CAS}
\defop{FAI}
\defop{FAD}
\defop{LL}
\defop{SC}
\defop{CL}
\defop{VL}
\defop{Write}
\defop{Swap}
\defop{cleanup}

\SetKwData{val}{val}
\SetKwData{Tag}{tag}

\title{Making History-Independence-Preserving LL/SC Swappable}
\author{Dante Bencivenga}{University of Calgary, Canada}{drbenciv@ucalgary.ca}{https://orcid.org/0000-0002-4481-7851}{Natural Sciences and Engineering Research Council of Canada (NSERC) Canada Graduate Scholarship -- Doctoral (CGS-D) 578918-2023/198 and University of Calgary Eyes High Doctoral Recruitment Scholarship}
\author{Philipp Woelfel}{University of Calgary, Canada}{woelfel@ucalgary.ca}{https://orcid.org/0000-0002-7847-4631}{Natural Sciences and Engineering Research Council of Canada (NSERC) reference number  RGPIN-2025-04233} %
\authorrunning{D. Bencivenga and P. Woelfel}
\Copyright{Dante Bencivenga and Philipp Woelfel}

\begin{document}

\renewcommand*{\subsectionautorefname}{Section}
\renewcommand*{\subsubsectionautorefname}{Section}

\maketitle

\begin{abstract}
	A randomized implementation of $m$ LL/SC objects for $n$ processes, with at most $k$ outstanding \LL{} operations each, uses $O(kn + m)$ CAS and FAID objects~\cite{BHW2026a}, achieves constant expected step complexity, and preserves quiescent history independence for algorithms such as~\cite{ABFOS2025a}.
	Unlike some other LL/SC algorithms, such as~\cite{AM1995a,JP2003a,JJJ2023a,ABFOS2024a}, this algorithm does not support an unconditional write operation.
	We show how to add a (randomized) wait-free swap operation, and therefore also a write operation, without affecting the asymptotic step or space complexity of the implementation.
\end{abstract}

\section{Introduction}\label{sec:intro}

Asynchronous shared memory algorithms generally require the use of atomic hardware instructions in order to coordinate their actions.
Several such instructions are now standard in modern processors, such as swap, compare-and-swap, and fetch-and-add:
The swap instruction writes a value into a shared memory word and returns its immediately preceding value.
The compare-and-swap, or \CAS{$old, new$}, instruction only takes effect if the shared memory word stores an expected $old$ value, in which case it replaces it with $new$.
In this paper, we only need \CAS{} to return a Boolean indicating success or failure, but it can sometimes return the previous value like a \Swap{} call.
The fetch-and-add instruction adds a given integer value to the value stored in a shared memory word, returning the previous value.
In this paper, we need only use fetch-and-add to add or subtract one, which we call fetch-and-increment (\FAI{}) and fetch-and-decrement (\FAD{}) respectively.

Although \CAS{} together with read/write registers is already sufficient to implement any shared object with a sequential specification~\cite{Her1991a}, it can be difficult to use correctly due to the \emph{ABA problem}.
For example, suppose a shared memory object stores value $v_1$, is read by a process, and then the process attempts a \CAS{} from $v_1$ to $v_2$.
The process has no way of knowing whether the object's value stayed at $v_1$ in between the read and \CAS{}, or if it changed to $v_3$ and back in between.
This ambiguous situation is the ABA problem, and it can easily lead to incorrect changes to data structures if not handled carefully.

The \emph{LL/SC} object fixes this problem by the use of two operations.
The load-linked (\LL{}) operation reads and returns the object's value.
The store-conditional (\SC{v}) operation can \emph{succeed} or \emph{fail}:
It succeeds and writes a value $v$ to the object if and only if no \Swap{} or successful \SC{} took place since the preceding \LL{}.
Otherwise, it fails (see \autoref{sec:write:seq_spec} for a more formal specification).
Therefore, it prevents the ABA problem, hence avoiding a large class of bugs that are otherwise hard to identify and correct.
Most implementations, including ours, also include a verify-link (\VL{}) operation, which returns a Boolean indicating whether an \SC{} would succeed, without changing the state of the object.

Unfortunately, LL/SC as defined here does not exist in modern hardware.
(Some architectures have a pair of instructions called \LL{} and \SC{}, but their \SC{} operation can spuriously fail even without any change to the memory word, and the pair is strongly restricted in which operations can happen in between.
These instructions are sometimes called RLL/RSC to differentiate them from `strong' LL/SC; see, e.g.,~\cite{JP2003a}).
As such, several software implementations have been devised~\cite{IR1994a,AM1995b,Moi1997b,LMS2003a,DHLM2004,Mic2004,JP2005a,JP2005b,JP2005c,AW2016a,BW2020a,NW2024a,BHW2026a}, with various properties, such as supporting a \Write{} operation~\cite{AM1995a,Jay1998a,JP2003a,JJJ2023a,ABFOS2024a} (some of which can be trivially converted into \Swap{}~\cite{AM1995a,ABFOS2024a}), achieving strong linearizability~\cite{NW2024a}, durability~\cite{JJJ2023a}, or adapting to the number of processes in the system~\cite{JP2005c,DHLM2004,JJJ2023a}.
Although any LL/SC implementation can support a lock-free \Swap{$v$} operation by repeatedly calling \LL{} and then \SC{$v$} until some \SC{$v$} attempt succeeds, supporting wait-free \Swap{} and LL/SC on the same object is more difficult.

Our implementation has optimal space and expected step complexity, uses bounded base objects, supports a \Swap{} operation in addition to \LL{} and \SC{}, \emph{and} preserves quiescent history independence, which we explain next.

\emph{History independence} is a useful security property which has been heavily studied in the context of sequential computing~\cite{Mic1997a,NT2001a,HHMPR2002a,BP2006a,ABHVW2004a,HHMPR2005a,BG2007a,NSW2008a,Gol2009a,Gol2010a,BBJKMPSZ2016a} and recently explored in concurrent computing~\cite{SB2014a,ABFOS2024a,ABFOS2025a,BHW2026a}.
Roughly, an implementation is history-independent if the state of memory in between operations is a function only of the abstract state of the implemented object(s).
In the context of concurrent computing, there are several ways to define history independence, as explored in~\cite{ABFOS2024a}.
Their most stringent definition is \emph{perfect} history independence, which requires that shared memory is a function of the abstract state at \emph{all} points of the execution, but they show that this requires impractically large base objects for most abstract objects.

As with~\cite{BHW2026a}, we instead focus on \emph{quiescent history independence} (QHI).
A \emph{quiescent state} is a system configuration in which no process is \emph{pending}, i.e., no process has invoked but not yet responded from a method call.
An implementation is QHI if its state of shared memory is a function of the abstract state of the concurrent object(s) during all quiescent states reachable in any execution of that implementation.
Since algorithms using LL/SC generally do not require outstanding \LL{} operations (i.e., an \LL{} operation not yet followed by a matching \SC{} operation, which we formally define in \autoref{def:write:outstanding_ll}) in quiescent configurations, we use a relaxed definition of QHI, following~\cite{BHW2026a}:
We replace quiescent states in the definition of QHI with \emph{settled states}, which are quiescent states with the additional property that no process has an outstanding \LL{} operation.
The state of shared memory then needs to be a function of the abstract objects' state during settled states of any reachable execution.
An LL/SC implementation satisfying this relaxed definition, such as ours, is \emph{QHI-preserving}.
Replacing atomic LL/SC objects in a QHI algorithm by a QHI-preserving implementation preserves QHI, provided that the algorithm has no outstanding \LL{} operations when its processes are idle.
This is usually possible to achieve by adding ``clear-link'' or \CL{} operations before method responses, which clears outstanding \LL{} operations.

There are few LL/SC implementations which also provide a \Write{} operation with $o(m \cdot n)$ space complexity, where $m$ is the number of LL/SC objects and $n$ is the number of processes.
Jayanti, Jayanti, and Jayanti's implementation~\cite{JJJ2023a} has constant step complexity, optimal adaptive space complexity of $O(m + n + C)$, where $C$ is the total number of outstanding \LL{} operations at a given time, and it is durable on non-volatile memory.
However, it uses unbounded sequence numbers, and is therefore not suitable for history independent algorithms.
The implementation of Attiya, Bender, Farach-Colton, Oshman, and Schiller~\cite{ABFOS2024a} satisfies perfect history independence, but uses large base objects of size $O(n)$ bits and is only lock-free.

Other papers give transformations to add a \Write{} operation to existing LL/SC implementations:
Jayanti~\cite{Jay1998a} gave a transformation to obtain a writable LL/SC object from two non-writable LL/SC objects, attaching constant-size tags to the objects values. %
However, because the algorithm's correctness requires these tags to not be reset in quiescent states, the transformation does not preserve history independence.
Another, more general, transformation by Aghazadeh, Golab, and Woelfel adds unconditional \Write{} operations to shared objects~\cite{AGW2014a}, and Ben-David, Blelloch, Friedman, and Wei later optimized their space complexity~\cite{BBFW2019a}, but their transformation increases the space by $O(n^2)$ additional base objects.
Moreover, their transformation also does not preserve history independence, unlike our implementation.

Our implementation is based on that of Bencivenga, Habashi, and Woelfel~\cite{BHW2026a}, which has the above features but without a \Write{} operation.
The addition of a randomized wait-free \Swap{} operation (which can be used as a \Write{} operation by discarding the return value) and a new correctness analysis are the main technical contributions of this paper.
\begin{theorem}\label{thm:write:main}
  Let $n$ be the number of processes and $k$ be the maximum number of simultaneously outstanding \LL{} operations per process.
  Then there is a linearizable and QHI-preserving implementation of $m$ LL/SC objects from $m$ CAS/Swap objects, storing LL/SC values plus tags of $\log (kn) + O(1)$ bits, and $O(m + kn)$ FAID/CAS objects of $\log n + O(1)$ bits.
  The implementation supports the implemented methods \LL{}, \SC{}, \Swap{}, \CL{}, \Read{}, and \VL{}, with $O(1)$ expected step complexity under the weak adaptive adversary.
\end{theorem}
Because our implementation uses more base objects, larger tags, and more steps per implemented method than~\cite{BHW2026a}, while also using a wider variety of primitive operations, their implementation is preferable when unconditionally swapping or writing to LL/SC objects is not required.

Like the implementation of~\cite{BHW2026a}, we attach bounded tags to each CAS object that implements an LL/SC word, of size $\ceil{\log (k + 1) n} + 2$ bits.
As~\cite{BW2020a,BHW2026a} argue, $k$ is at most $3$ in published algorithms using LL/SC, thus making our tags of at most $\ceil{\log n} + 4$ bits in practice.
Since $64$-bit heap pointers generally have the high $16$ bits and low $3$ bits set to zero~\cite{Cha2022a}, this allows a pointer to be stored with $19$ tag bits into one $64$-bit word.
As a result, our implementation can be used to store pointer values in LL/SC objects if the number of processes is up to $2^{15} = 32{,}768$, which matches the default maximum number of processes and threads on Linux~\cite{Lin2026a}.

The additional complexity needed to support unconditional wait-free \Swap{} arises in part because it becomes more difficult to correctly choose a new tag to prevent ABAs.
In particular,~\cite{BHW2026a} counts how many processes have an outstanding \LL{} operation which have seen each tag value in FAID objects, and \SC{} randomly probes these usage counts to find a tag with a usage count of zero.
This works in part because a successful \SC{} must be preceded by an \LL{} operation, and therefore the previous tag has a nonzero usage count, ensuring that a successful \SC{} always changes the tag.
However, an unconditional \Swap{} operation has no such prerequisite, so a further mechanism is needed to ensure that it always changes that tag, so that concurrent \SC{} operations can fail.
As such, our usage counts additionally count how many LL/SC objects are currently using a given tag.
\SC{} and \Swap{} calls attempt a \CAS{} from zero to one on a randomly probed usage count, and then keep the usage count incremented until the LL/SC object no longer stores that tag.
This makes tracking usage counts more complicated, since the responsibility for decrementing them can shift to a different process, and it affects several other aspects of the analysis.

The remainder of the paper is structured as follows: \autoref{sec:prelims} defines the shared-memory model and the specification of swappable LL/SC, \autoref{sec:impl} presents our implementation and intuitive explanations for it,
\autoref{sec:analysis} proves \autoref{thm:write:main}, and
\autoref{sec:conclusion} concludes the paper.

\section{Preliminaries}\label{sec:prelims}

We use the standard asynchronous shared memory model, where $n$ processes communicate by performing atomic shared-memory operations on shared objects.
At each point in time of the execution (which we refer to simply as a `point'), exactly one shared memory operation takes place.

\subsection{Primitive Operations}\label{sec:write:primitives}
The atomic primitive operations we use are \Read{}, \CAS{}, \FAI{}, \FAD{}, and \Swap{}.
The \Read{} operation returns the value of the shared object.
The \Swap{val} operation writes $val$ into the shared object, and returns its previous value.
The \CAS{old, new} operation writes $new$ into the shared object and returns $\True$ if and only if its previous value was $old$, and otherwise leaves the object unchanged and returns $\False$.
Finally, the \FAI{} and \FAD{} increment or decrement the object's (integer) value, respectively, and returns the previous value.
In our analysis, we initially assume that a \FAI{} on value $max$ wraps to back to $0$, and similarly that a \FAD{} on $0$ wraps it to $max$.
We prove (\autoref{lem:write:C_value_equals_contributions} and \autoref{lem:write:R_value_and_mutex_lock}(\ref{item:write:R_value_equals_contributions})) that overflow and underflow never occur on the FAID objects, and therefore those objects can have any overflow and underflow behaviour (including having a larger range of possible values before overflowing).

The objects in our implementation require the combinations \Read{}, \CAS{}, and \Swap{} for elements of an array $L$, \Read{}, \FAI{}, and \FAD{} for elements of an array $C$, and \FAI{}, \FAD{}, and \CAS{0, 1} for elements of an array $R$.
(In practical implementations, all shared memory variables generally support the same set of atomic hardware instructions, which commonly include all the ones used in this paper.)

\subsection{Adversary}
Our algorithm is randomized, meaning that at any point a process can generate a (private) random number by performing a \emph{random choice step}.
The adversary determines the order in which processes take steps, by deciding at any point during the execution, the next process to take a step.
We assume the same adversary used in~\cite{BHW2026a}, the \emph{weak adaptive adversary}, where a process always performs its next shared memory operation immediately after a random choice step.
I.e., while the weak adaptive adversary knows all past random choices and adapts to them, it cannot intervene between a process's random choice and its following shared-memory step.
Even though we employ randomization, all correctness properties such as linearizability~\cite{HW1990a} and preserving quiescent history independence (QHI) are achieved deterministically, i.e., they hold for all executions, regardless of the random choices made.

\subsection{Sequential Specification}\label{sec:write:seq_spec}
This section closely follows Section~2.1 of~\cite{BHW2026a}, with all operations other than \Swap{} defined in the same way as in that paper.

An \emph{abstract object} (also known as \emph{type}) is a quintuple $(Q,q_0,I,R,\delta)$, where $Q$ is a set of states, $q_0$ is the initial state, $I$ and $R$ are sets of operations and responses, respectively, and $\delta:Q\times I\to Q\times R$ is the transition function.
This quintuple is also called the \emph{sequential specification} of the object, as it defines the set of possible sequential executions: Whenever an object is in state $q$ and operation $i$ is invoked, $\delta(q,i)=(q',r)$ defines the operation's response $r$ and resulting new state $q'$.

To specify the abstract object swappable LL/SC, we follow the definitions in~\cite{JJJ2023a,ABFOS2024a,BHW2026a}.
Each state is a pair $(v,\mathcal{L})$, where $v$ is the object's \emph{value} and $\mathcal{L}$ is an $n$-bit sequence $(\ell_1,\dots,\ell_n)$, called \emph{context}.
Initially, $\ell_1=\cdots=\ell_n=0$, and the initial value of $v$ can be defined arbitrarily.
We say that $p$ has an \emph{open link} to the LL/SC object whenever $\ell_p = 1$.

A \Read{} returns the value $v$ of the object.
An \LL{} (``load-linked'') call by process $p$ sets $\ell_p$ to 1 and returns the value $v$.
A \Swap{x} operation by process $p$ unconditionally changes the state of the object to $(x,(0,\dots,0))$ and returns its previous value $v$.
(A \Write{x} operation is the same as \Swap{x} except that it does not return anything, and it can therefore be implemented from \Swap{x}.)
An \SC{x} (``store-conditional'') operation by process $p$ succeeds if $\ell_p=1$, in which case it changes the state of the object to $(x,(0,\dots,0))$ (like \Swap{x}) and returns $\True$.
If $\ell_p=0$, then the operation fails and returns $\False$ without changing the state of the object.
Operation \CL{} (``clear link'') by process $p$ simply resets $\ell_p$ to 0, and does not return anything.
Finally, operation \VL{} (``verify link'') by process $p$ returns a Boolean value that is true if and only if $\ell_p=1$.

In this paper, we implement a collection of $m$ swappable LL/SC objects, and refer to the $i$-th such object as $L_i$.
Thus, the abstract object's state is the cross product of the states of $m$ individual LL/SC objects.
As such, each operation has one additional parameter that indicates which of the LL/SC objects in the collection is affected.
For example, operation \Swap{i,x} behaves as the \Swap{x} operation defined above, affecting only $L_i$, the $i$-th LL/SC object in the collection.

\subsection{Outstanding LL() Operations}\label{sec:write:pending_links}
Our implementation requires knowledge of the number of processes and the number of outstanding \LL{} operations per process, which we formally define next:
\begin{definition}[Outstanding \LL{} Operation]\label{def:write:outstanding_ll}
  A process $p$ has an outstanding \LL{} operation on $L_i$ at point $t$ if and only if
  \begin{itemize}
    \item $p$ has no pending method call on $L_i$ at point $t$, and
    \item $p$ has called \LL{i} before point $t$ without having subsequently invoked a \Swap{i, \cdot}, \SC{i, \cdot}, or \CL{i} call, or a \VL{i} call that returns \False.
  \end{itemize}
  Equivalently, we say that $p$ has an outstanding \LL{i} at point $t$.
\end{definition}
The definition of outstanding \LL{} operations is similar to that of~\cite{BHW2026a}, except that calls to \Swap{i, \cdot} clears any outstanding \LL{i}, and we do not consider a process with a pending call on $L_i$ to have an outstanding \LL{i}.

In our implementation, $k$ is a per-process bound on the number of outstanding \LL{} operations (as opposed to a global bound, as in~\cite{BHW2026a}).
As such, we require that when \LL{i} is called, the calling process has fewer than $k$ outstanding \LL{} operations to objects other than $L_i$.

As with~\cite{BHW2026a}, our implementation requires that processes clear their outstanding \LL{i} operations with \CL{i} if they are no longer needed before returning from methods, in order to achieve our space bounds and preserve QHI\@.

\section{Implementation}\label{sec:impl}

\subsection{Intuition for Changes from~\texorpdfstring{\cite{BHW2026a}}{Non-Writable QHI-Preserving LL/SC}}\label{sec:changes}
At a very high level, the non-writable LL/SC implementation of~\cite{BHW2026a} works by storing the value of the $i$-th LL/SC object, $L_i$, together with a non-negative bounded tag in an array entry $L[i]$.

An array $R$ has one element per nonzero tag $g$: $R[g]$ stores the \emph{usage count} of tag $g$, which is the number of outstanding \LL{} operations that have read tag $g$ from the \Tag field of some element of $L$.
(Tag zero is the default value, and as such is treated differently, including not needing a usage count.)
An \LL{i} call increments $R[g]$ if it reads that $L[i].\Tag = g \neq 0$.
A nonzero usage count signals that a process has \emph{protected} that tag, and that it should not be reused while protected.
An \SC{i, \cdot} call randomly probes elements of $R$ until it finds a new tag $g'$ such that it reads $R[g'] = 0$.
A usage count of zero indicates that $g'$ is not protected, and it is hence safe to change $L[i].\Tag$ to $g'$ without causing an ABA problem.
Any operation that clears an outstanding \LL{} operation then decrements the usage count of its associated tag.

Another array $C$ has one element per LL/SC object: $C[i]$ stores the \emph{activity count} of $L_i$, which is the number of processes with a pending method call on $L_i$ or that have an outstanding \LL{i}.
When $C[i]$ reaches zero, it is safe (and necessary for preserving QHI) to reset $L[i].\Tag$ to zero, since there is no outstanding \LL{i} and therefore no ABA problem to guard against.
A nonzero activity count can thus be roughly thought of as protecting tag zero from being used in non-quiescent states.

For more thorough explanations, see the high- and low-level descriptions in Section~3 of~\cite{BHW2026a}.

We now explain the motivation for how to change the algorithm of~\cite{BHW2026a} to allow for randomized wait-free \Swap{} operations, and the necessary changes required to maintain correctness, the same asymptotic step and space complexity, and to preserve quiescent history independence.

The primary difference in our implementation from that of~\cite{BHW2026a} is in how we manage the usage counts in array $R$.
As with~\cite{BHW2026a}, when a process performs \LL{i} on LL/SC object $L_i$ and sees a nonzero tag $g$, it increments usage count $R[g]$.
(The default tag zero does not have a usage count, and its ABA prevention comes from activity counts in array $C$).

Recall that in~\cite{BHW2026a}, an \SC{i, \cdot} call randomly probes elements of $R$ until it reads zero for some $R[g']$---this implies that, at the time of the read, tag $g'$ is not protected.
However, it is possible that tag $g'$ becomes protected by the time the \SC{} call attempts a \CAS{} on $L[i]$, which if successful would change $L[i].\Tag$ to the protected tag $g'$, thus potentially causing an ABA problem.
But, a process only protects tag $g'$ if it sees that $L[i].\Tag = g'$, implying that $L[i]$ changes before the attempted \CAS{} on $L[i]$.
Moreover, the process calling \SC{i, \cdot} is itself protecting a tag $g$ from its own outstanding \LL{i} that it read from $L[i].\Tag$.
It uses tag $g$ in its CAS attempt, i.e, the \CAS{} is from $(\cdot, g)$ to $(\cdot, g')$.
Thus, the only way for the \CAS{} to succeed is if $L[i].\Tag$ changes from $g$ to $g'$ and back to $g$, i.e., by an earlier ABA\@.
Since \SC{} is the only method in their implementation which can modify $L[i]$, and since it cannot cause the \emph{first} ABA, it follows that it cannot cause \emph{any} ABA\@ (see their Lemma 17).
Crucially, their argument relies on only \SC{} calls modifying $L[i]$, and that $L[i]$ is only modified via \CAS{} operations.

Like \SC{i, \cdot}, an added \Swap{i} operation should also choose a new tag that no processes' currently outstanding \LL{i} has seen, in order to prevent ABAs for those processes' \SC{i, \cdot} operations.
Unlike \SC{}, however, \Swap{} always succeeds and does not require a preceding \LL{}, while still being (randomized) wait-free.
This means that simply verifying that a usage count is zero is no longer sufficient:
A process $p$ executing a \Swap{i, \cdot} operation could fall asleep after reading $R[g] = 0$, after which tag $g$ could become used in $L[i].\Tag$ from another \SC{i, \cdot} or \Swap{i, \cdot} operation.
Since we require that \Swap{} is a randomized wait-free method which unconditionally updates $L_i$, there must be some step in its operation in which it updates $L[i]$ with some nonzero probability, along with updating $L[i].\Tag$ to a new, unprotected tag.
But since a \Swap{} call does not know that $g$ is still unprotected before its update to $L[i]$ after having only read that $R[g] = 0$, any such update to $L[i]$ can potentially lead to ABAs.
(In contrast, an \SC{} attempts a \CAS{}, and this is guaranteed to fail if $L[i]$ had changed, as argued earlier.)

Instead, our approach is for $p$ to prevent its chosen tag $g$ from being used by any concurrent \Swap{} or \SC{} operation by replacing a read of zero with an $R[g].\CAS{0, 1}$ operation.
This effectively \emph{locks} the tag for exclusive use by $p$ until $R[g]$ is eventually decremented back to zero.
Following~\cite{BHW2026a}, we assume that processes are scheduled by the \emph{weak adaptive adversary} which cannot intervene in between a random choice and the next shared-memory step.
Thus, after a random choice of a new tag $g$, we immediately follow it by an $R[g].\CAS{0, 1}$ attempt, which succeeds with constant probability as long as we have sufficiently many unused tags to choose from.
We call such a successful \CAS{} on $R[g]$ \emph{locking} tag $g$ by the calling process (this lock does not compromise randomized wait-freedom, since our algorithm guarantees a constant success probability with each randomly probed tag).
A \Swap{i, u} call can then unconditionally write this new tag $g$ and its value $u$ into $L[i]$.
We also need to prevent \Swap{} calls from using the same tag chosen by concurrent \SC{} calls, which can cause ABA's due to the unconditional nature of \Swap{}:
We therefore change the \SC{} method to attempt a \CAS{0, 1} for each randomly selected tag until it succeeds, exactly like the \Swap{} method.

In order to ensure history independence and allow for eventual safe recycling of tags, this additional increment of $R[g]$ must eventually be decremented before the system enters a quiescent state.
In the case of a failed \SC{} operation, the calling process can safely decrement the usage count for its selected tag immediately, since it is not used anywhere else.

However, consider what happens if we decrement $R[g]$ at the end of a \Swap{i, \cdot} or successful \SC{i, \cdot} call.
If some other crashed process keeps $C[i] > 0$, indicating a non-quiescent state, then $L[i].\Tag$ is not reset to zero, in which case we could have $L[i].\Tag = g$ and $R[g] = 0$ when a \Swap{} or \SC{} call responds.
This scenario is not a problem in~\cite{BHW2026a}, when only \SC{} operations can change LL/SC objects.
In our case, however, a new \Swap{} call could re-lock tag $g$, reaching a configuration with a pending \Swap{} operation poised to write $g$ back into $L[i].\Tag$, while $L[i].\Tag = g$.
This scenario violates the sequential specification, because \Swap{i, \cdot} should set all context bits of $L_i$ to zero, causing concurrent \SC{i, \cdot} operations to fail; writing back the same value-tag pair is instead invisible to all other processes.
Also, a \Swap{i, \cdot} call poised to overwrite $L[i]$ could revert a changed value-tag pair, which again leads to an undetected ABA, invisible to concurrent \LL{i} operations regardless of which tags they protect.

While it is tempting to have \Swap{} calls simply double-check that $L[i].\Tag \neq g$ after locking $g$, this would in the worst case require a lock-free loop that cannot guarantee wait-freedom.
(Recall that we only prevent the weak adaptive adversary from interfering in between a random choice and the immediate next step, whereas this last idea would require two shared-memory steps to be uninterrupted to achieve randomized wait-freedom.)

Due to these considerations, after locking a nonzero tag $g$, we keep it locked for at least as long as $L[i].\Tag = g$.
This implies that the responsibility to decrement $R[g]$ shifts from the process with the \Swap{} or successful \SC{} operation to the one that next changes $L[i].\Tag$ to some other tag.
After every successful \CAS{} operation on $L[i]$ which changes $L[i].\Tag$ away from $g$, we thus add a decrement of $R[g]$.
Since the implemented \Swap{} operation needs to unconditionally change $L[i]$ and also know the previous value and tag stored in $L[i]$ before overwriting it, we use the \Swap{} hardware instruction to read the previous value and tag.
(The need to read the previous tag in particular is part of the motivation for why we implement the stronger \Swap{} operation, and not just \Write{}.)
Thus, the \Swap{} method also decrements $R[g]$ after reading nonzero $g$ from $L[i].\Tag$ after a hardware \Swap{}, without losing randomized wait-freedom.

Since \Swap{} and \SC{} now increment usage counts, and the responsibility for decrementing the counts shifts to other processes, analyzing contributions to usage counts becomes more complicated.
A side effect of keeping tags locked by LL/SC objects is that each object is now guaranteed to store a unique tag if it is nonzero (i.e., distinct LL/SC objects do not share the same nonzero tag).
This property is not directly needed for correctness, complexity, or preserving QHI, but it does reduce the maximum value that elements of $R$ need to store to $n + 1$, compared to $kn$ in~\cite{BHW2026a}.

Because objects now contribute to usage counts, it may initially appear that we need $\Omega(m)$ more tags, and hence $\Omega(m)$ more elements of $R$, so that the choice of a random tag still succeeds with constant probability.
However, recall that, to preserve QHI, we reset all \Tag fields to zero for each object that does not have a pending method call or an outstanding \LL{} operation on it.
As such, we can show (\autoref{lem:write:max_total_contribution}) that every LL/SC object $L_i$ with a nonzero \Tag field is associated with some process that either has a pending method call on $L_i$ or has an outstanding \LL{i}.
Since there are $n$ processes and at most $kn$ outstanding \LL{} operations, the extra number of used tags is at most $(k + 1)n$.
In total, including the factor of $2$ to ensure that at least half of the tags are unused, $4(k + 1)n$ nonzero tags, only a constant factor more than the $2kn$ nonzero tags required in~\cite{BHW2026a}.

This observation only holds when using the QHI-preserving version of the algorithm: as with~\cite{BHW2026a}, our implementation can be made simpler by dropping that requirement, but then the number of required tags becomes $2((k + 1)n + m)$.
Conversely, if \Swap{i, \cdot} is only invoked when the calling process has fewer than $k$ outstanding \LL{} operations, then this bound decreases slightly to $(4k + 2)n$.
Regardless, all such variants of the implementation use $\Theta(kn + m)$ total base objects.

Lastly, to preserve QHI, the \Swap{} method must change the \Tag field to zero if the activity counter is zero, using the same mechanism as in~\cite{BHW2026a} in their \CL{} method.
However, since the \Swap{} method does not involve any outstanding \LL{} operation, it does not need to do the bookkeeping work of clearing information related to outstanding \LL{} operations like in that paper's \CL{} method.
Since the \Swap{} method also does not guard against an ABA problem, it has no need to increment (or decrement) the activity counter itself:
It only needs to check whether the counter is zero and, if so, to try to change the LL/SC object's \Tag field to zero.
As such, we split \CL{} into a shorter \CL{} method plus a \cleanup{} method: the former handles clearing outstanding \LL{} operations, and the latter attempts to reset \Tag fields to zero before entering a quiescent state.

\subsection{Low-Level Description}

\newcommand{\HI}[1]{{\color{black}#1}}
\newcommand{\W}[1]{{\color{blue}#1}}
\newcommand{\WHI}[1]{{\color{blue}#1}}
\newcommand{\nrm}[1]{{\color{black}#1}}
\newcommand{\lHI}{\textcolor{black}}
\newcommand{\lnrm}{\textcolor{black}}

In this section, we present and explain the pseudocode for our implementation, including the parts that are unchanged from~\cite{BHW2026a}.
When the details are the same, we explain them similarly.

We store the values of the LL/SC objects in an array $L$ of CAS objects.
In addition, each CAS object stores a non-negative bounded tag that is initialized to zero.
The array $R$ is used to count each tag's usage, so that $R[g]$ counts the number of processes that have an outstanding \LL{} operation that read the tag $g$, plus the number of objects with $g$ in their \Tag field.
The purpose of usage counts is to prevent ABAs.
Array $C$ is an activity counter for LL/SC objects, so that $C[i]$ tracks how many processes have an outstanding \LL{i}; its purpose is to detect quiescent states, and thus help preserve quiescent history independence.

\begin{figure}
  \begin{invisiblebox}{\KwSty{Constants:}}
    \begin{itemize}\fontsize{9.5}{11.5}\selectfont
      \item $U$ is the set of values that LL/SC objects can have
      \item $m$ is the number of LL/SC objects
      \item $n$ is the number of processes
      \item $k$ is the maximum number of outstanding \LL{} operations per process
    \end{itemize}
  \end{invisiblebox}

  \begin{invisiblebox}{\KwSty{Shared Data:}}
    \begin{itemize}\fontsize{9.5}{11.5}\selectfont
      \item $L[i]$ for $i \in \set{1, \dots, m}$ is a CAS\W{/Swap} object storing a pair in\\ $U \times \set{\HI{0,\,}1, \dots, \W{4(k + 1)n}}$
      \item \HI{$C[i]$ for $i \in \set{1, \dots, m}$ is a FAID object storing a number in $\set{0, \dots, n}$}
      \item $R[g]$ for $g \in \set{1, \dots, \W{4(k + 1)n}}$ is a FAID\W{/CAS} object storing a number in $\set{0, \dots, \W{n + 1}}$
    \end{itemize}
  \end{invisiblebox}

  \begin{invisiblebox}{\KwSty{Process-local static data:}}
    \begin{itemize}\fontsize{9.5}{11.5}\selectfont
      \item $d$ is a dictionary storing keys in $\set{1, \dots, m}$ and values in\\ $U \times \set{0, 1, \dots, \W{4(k + 1)n}}$
    \end{itemize}
  \end{invisiblebox}

  \begin{invisiblebox}{\KwSty{Initialization:}}
    \begin{itemize}\fontsize{9.5}{11.5}\selectfont
      \item $L[i] \gets (u_i, \HI{0})$, where $u_i$ is an application-dependent initial value from $U$
      \item \HI{$C[i] \gets 0$}
      \item $R[g] \gets 0$
      \item $d$ starts empty for each process (all indexing returns $\bot$)
    \end{itemize}
  \end{invisiblebox}
\end{figure}

In any LL/SC implementation, processes need some way to keep track of their outstanding \LL{} operations, in order to be able to know whether their \SC{} and \VL{} calls should succeed.
Here, processes use a local dictionary $d$, which stores, for each LL/SC object $L_i$, a pair consisting of the last value and tag read from $L[i]$, or the empty value $\bot$ if the process has no outstanding \LL{i}.
(The \SC{i, \cdot}, \VL{i}, and \CL{i} methods each return immediately when $d[i] = \bot$, since they are only meaningful for a process with an outstanding \LL{i}.)

Changes or additions relative to~\cite{BHW2026a} are indicated in \W{blue}.
Our implementation uses $4(k + 1)n$ tags rather than $2kn$ as in~\cite{BHW2026a}, or slightly over twice as many tags.
On the other hand, in this version tags are never shared between distinct LL/SC objects, the elements in $R$ need only store values up to $n + 1$ rather than $kn$.
In addition to adding a \Swap{} method relative to~\cite{BHW2026a}, we also move the \Tag field-resetting functionality of \CL{} into a new helper method called \cleanup{}.
Since \cleanup{} is only a helper method, it is not called directly, and as such is not mentioned in the sequential specification.

\subsubsection{LL() Method}

The \LL{} method is unchanged from that of~\cite{BHW2026a}.

\begin{figure}
    \begin{method}{LL($i \in \set{1, \dots, m}$)}
    \lIf{$d[i] \neq \bot$}{\CL{i}}\label{line:ll:reset_link}
    \HI{$C[i].\FAI{}$}\;\label{line:ll:inc_activity}
    $(val, tag) \gets L[i].\Read{}$\;\label{line:ll:first_read}
    $d[i] \gets (val, tag)$\;\label{line:ll:set_di}
    \HI{\lIf{$tag = 0$}{\Return{$val$}}}\label{line:ll:first_zero_tag}
    $R[tag].\FAI{}$\;\label{line:ll:inc_refcount}
    $(val', tag') \gets L[i].\Read{}$\;\label{line:ll:second_read}
    \lIf{$(val, tag) = (val', tag')$}{\Return{$val'$}}\label{line:ll:matching_tag}
    \HI{\If{$tag' = 0$}{
      $R[tag].\FAD{}$\;\label{line:ll:dec_refcount}
      $d[i] \gets (val', 0)$\;\label{line:ll:reset_tag}
      \Return{$val'$}\label{line:ll:second_zero_tag}
    }}
    \CL{i}\;\label{line:ll:clear}
    \Return{$val$}\label{line:ll:final_return}
  \end{method}
\end{figure}

Since processes can have at most one outstanding \LL{i}, \autoref{line:ll:reset_link} ensures that any previous outstanding \LL{i} is cleared by an internal \CL{i} call (\autoref{line:swap:clear} of \Swap{} serves a similar purpose).
The calling process $p$ then increments $C[i]$ in \autoref{line:ll:inc_activity}, which announces that a process has created an outstanding \LL{i}, and tells processes not to reset $L[i].\Tag$ to zero.
Next, $p$ reads the value-tag pair stored in $L[i]$ (\autoref{line:ll:first_read}), and stores it in its local dictionary $d$ (\autoref{line:ll:set_di}), to use in later \SC{i, \cdot} or \VL{i} calls.
If the tag is the default tag of zero, then the call can respond immediately (\autoref{line:ll:first_zero_tag}), because the \SC{} and \Swap{} methods do not choose tag zero as their new tag, and a nonzero $C[i]$ prevents ABAs to tag zero.
Otherwise, $p$ reads a nonzero tag $g$ and increments the corresponding usage count $R[g]$ (\autoref{line:ll:inc_refcount}), then reads $L[i]$ again to confirm that it has not changed (\autoref{line:ll:second_read}).

If $L[i]$ has not changed upon the second read, then we say that $p$ has successfully \emph{protected} the tag from being reused (see the formal definition in \autoref{def:write:protect}), and $p$ returns in \autoref{line:ll:matching_tag}.
This protection lasts until $p$ next invokes \CL{i}, as \autoref{line:cl2:dec_refcount} decrements the usage count again.
As we later discuss in the \SC{} method, the algorithm guarantees that a tag is only reused when no process protects it.
In this case, the method's linearization point (which defines the order of the execution's linearization) occurs at \autoref{line:ll:second_read}.

Otherwise, $L[i]$ \emph{has} changed to a different value by the time $p$ reads it in \autoref{line:ll:second_read}.
If $L[i].\Tag$ has changed to zero, then $p$ can return once it has reverted its increment of $R[g]$ (\autoref{line:ll:dec_refcount}) and updated it dictionary (\autoref{line:ll:reset_tag}), again because tag zero is protected by a nonzero activity count in $C[i]$.
Otherwise, we take advantage of the fact that an \SC{i, \cdot} is required to fail if there was an intervening \Swap{i, \cdot} or \SC{i, \cdot} in between a process's \LL{i} and \SC{i, \cdot}.
We thus retroactively set $p$'s \LL{i} call's linearization point to the first read in \autoref{line:ll:first_read} to satisfy the sequential specification.
(Although this is a standard trick for LL/SC implementations, this retroactive linearization point prevents this \LL{} implementation from being strongly linearizable~\cite{GHW2011a}.)
The \CL{i} call in \autoref{line:ll:clear} sets $d[i]$ to $\bot$ (\autoref{line:cl2:clear_di}), as though $p$ had no outstanding \LL{i}, which forces $p$'s following \SC{i, \cdot} to fail in \autoref{line:sc2:early_return}.
(In this case, $p$ still does have an outstanding \LL{i} by definition, because the fact that the next \SC{i, \cdot} will fail is not yet directly visible by an algorithm using our LL/SC implementation.)

\subsubsection{Swap() Method}

In lines~\ref{line:swap:repeat}--\ref{line:swap:lock} of a \Swap{i} call, the calling process $p$ repeatedly probes random tags in the range of possible tag values.
Specifically, \autoref{line:swap:choose_tag} chooses an independent and uniform random tag value in the domain of $4(k + 1)n$ possible tags.
The calling process attempts a \CAS{} from zero to one for the chosen tag's usage count, and repeats this random choice if the \CAS{} fails, until it eventually succeeds in incrementing some tag's usage count.
There are $4(k + 1)n$ tags, and as argued in the end of \autoref{sec:changes} and proven in \autoref{lem:write:max_total_contribution}, there are at most $2(k + 1)n$ tags with nonzero usage count at any given time.
Because the weak adaptive adversary cannot intervene between the random choice in \autoref{line:swap:choose_tag} and the \CAS{} attempt in \autoref{line:swap:lock}, each \CAS{} attempt succeeds with probability at least $1/2$.
Thus, the number of loop iterations is dominated by a geometrically distributed random variable with parameter $1/2$, which has expectation at most $2$.

\begin{figure}
  \begin{method}{\W{Swap($i \in \set{1, \dots, m}, u \in U$)}}
    \W{
      \lIf{$d[i] \neq \bot$}{\CL{i}}\label{line:swap:clear}
      \lnl{line:swap:repeat}\Repeat{
        $R[tag'].\CAS{0, 1} = \True$}
        {$tag' \gets$ \KwSty{independent uniform random sample from} $\set{1, \dots, 4(k + 1)n}$\label{line:swap:choose_tag}
      }\label{line:swap:lock}
      $(v, tag) \gets L[i]$.\Swap{(u, tag')}\;\label{line:swap:swap}
      \lHI{}\WHI{\lIf{$tag \neq 0$}{\W{$R[tag].\FAD{}$}}}\label{line:swap:dec_prev}
      \WHI{\lIf{$C[i].\Read{} = 0$}{\cleanup{i}}}\label{line:swap:confirm_quiescence}
      \Return{$v$}\label{line:swap:return}
    }
  \end{method}
\end{figure}

With a successful $R[g].\CAS{0, 1}$ for some nonzero tag $g$, $p$ \emph{locks} $g$ so that no other \Swap{} or \SC{} call can use it, and $g$ then remains locked for as long as it is used in $L[i].\Tag$.
Process $p$ then calls the \Swap{} hardware instruction on $L[i]$ in \autoref{line:swap:swap} to write its new value together with the locked tag $g$, which serves as the linearization point of the method.
Immediately after this point, we say that tag $g$ is locked by $L_i$, because $L[i].\Tag = g$.
Since $p$ had locked its nonzero tag $g$, and the previous value of $L[i].\Tag$ was either previously locked by $L_i$ or was zero, it is guaranteed that \autoref{line:swap:swap} changes $L[i].\Tag$ to a different value.
This in turn guarantees that any outstanding \LL{i} by other processes detects that the \Swap{} has linearized, preventing undetected ABAs.
Since the previous tag, read in \autoref{line:swap:swap}, is no longer used in $L[i].\Tag$, $p$ then decrements its usage count in \autoref{line:swap:dec_prev} so that it can eventually be \emph{unlocked} and used again once the usage count reaches zero.
(If there are outstanding \LL{i} operations by other processes which have seen tag $g$, then they contribute to the usage count $R[g]$, and so the tag is not fully unlocked until each of them have called \CL{i}.)

To preserve QHI, $p$ checks if the activity counter $C[i]$ is zero, in which case it is safe to reset the tag to zero via a \cleanup{i} call (\autoref{line:swap:confirm_quiescence}).
Note that \Swap{i} does not increment (or decrement) $C[i]$, because it changes $L[i]$ unconditionally unlike \SC{}, without needing ABA prevention for its own operation.

Lastly, \autoref{line:swap:return} returns the LL/SC value that was read at the hardware \Swap{} in \autoref{line:swap:swap}.

\subsubsection{SC() Method}\label{sec:low-level:sc}

Line~\ref{line:sc2:early_return} ensures that an \SC{} call returns \False in the case of the calling process not having any outstanding \LL{} (lines~\ref{line:cl2:early_return} of \CL{} and~\ref{line:vl:early_return} of \VL{} serve a similar purpose).
Lines~\ref{line:sc2:repeat}--\ref{line:sc2:lock} work just like lines~\ref{line:swap:repeat}--\ref{line:swap:lock} to \emph{lock} an unprotected tag for exclusive use by the calling process.
Recall that $d[i]$ stores the previous value-tag pair read from $L[i]$ in a previous \LL{i} operation.
Line~\ref{line:sc2:first_cas} thus attempts a \CAS{} from the previous value-tag pair stored in $d[i]$ to the new value (the second argument of \SC{}) and the newly chosen tag.
Similarly to \autoref{line:swap:dec_prev}, \autoref{line:sc2:dec_prev} unlocks the nonzero tag previously stored in $L[i]$ after a successful \CAS{} in \autoref{line:sc2:first_cas}, so that it can eventually be recycled.

\begin{figure}
  \begin{method}{SC($i \in \set{1, \dots, m}, u \in U$)}
    \lIf{$d[i] = \bot$}{\Return{\False}}\label{line:sc2:early_return}
    \lnl{line:sc2:repeat}\Repeat{
      $R[tag'].\W{\CAS{0, 1} = \True}$}
      {$tag' \gets$ \KwSty{independent uniform random sample from} $\set{1, \dots, \W{4(k + 1)n}}$\label{line:sc2:choose_tag}
    }\label{line:sc2:lock}
    $result \gets L[i]$.\CAS{d[i], (u, tag')}\;\label{line:sc2:first_cas}
    \W{\lIf{$result = \True$ \WHI{\KwSty{and} $d[i].\Tag \neq 0$}}{$R[d[i].\Tag].\FAD{}$}}\label{line:sc2:dec_prev}
    \HI{\lIf{$result = \False$}{$result \gets L[i]$.\CAS{(d[i].\val, 0), (u, tag')}}}\label{line:sc2:second_cas}
    \W{\lIf{$result = \False$}{$R[tag'].\FAD{}$\label{line:sc2:unlock}}}
    \CL{i}\;\label{line:sc2:clear}
    \Return{$result$}\label{line:sc2:return}

  \end{method}
\end{figure}

Although a nonzero activity counter ($C[i]$) prevents \CL{} calls from resetting $L[i].\Tag$ to zero (as we detail below), there is an edge case of a \CL{} call poised to clear $L[i].\Tag$ from before a time when $C[i]$ was incremented.
As such, \autoref{line:sc2:second_cas} tries another \CAS{} from a tag of zero, in case $L[i].\Tag$ was reset to zero in between the previous \LL{i} operation's linearization point and the \CAS{} attempt in \autoref{line:sc2:first_cas}.
If both \CAS{} attempts fail, then \autoref{line:sc2:unlock} unlocks the newly chosen tag, since it was not successfully used.
(In this case, the usage count immediately hits zero, since no other process would have seen the tag---\autoref{line:sc2:unlock} could be replaced with a \Write{} of zero, but this is somewhat less intuitive in the code.)

The method linearizes at the point of its last \CAS{} attempt on $L[i]$ in either \autoref{line:sc2:first_cas} or~\ref{line:sc2:second_cas}, then calls \CL{} in \autoref{line:sc2:clear} to clear the calling process's outstanding \LL{}, and returns whether its last \CAS{} attempt succeeded.

\subsubsection{CL() and cleanup() Methods}

After confirming that a \CL{i} call is nontrivial (\autoref{line:cl2:early_return}), the call ends the calling process $p$'s protection of a nonzero tag by decrementing that tag's usage count (\autoref{line:cl2:dec_refcount}) and clears $p$'s dictionary entry for $L_i$ (\autoref{line:cl2:clear_di}), ending $p$'s outstanding \LL{i}.
Next, $p$ decrements the activity counter for $L_i$ (\autoref{line:cl2:dec_activity}), announcing that its outstanding \LL{i} has been cleared.
If the counter decrements from one to zero, then $p$ has the responsibility to reset $L[i].\Tag$ to ensure QHI, which it does via the \cleanup{} method.

\begin{figure}
  \begin{method}{CL($i \in \set{1, \dots, m}$)}
    \lIf{$d[i] = \bot$}{\Return}\label{line:cl2:early_return}
    \lnrm{}\HI{\lIf{$d[i].\Tag \neq 0$}{\nrm{$R[d[i].\Tag].\FAD{}$}}}\label{line:cl2:dec_refcount}
    $d[i] \gets \bot$\;\label{line:cl2:clear_di}
    \HI{\lIf{$C[i].\FAD{}\ \W{=}\ 1$}{\W{\cleanup{i}}}}\label{line:cl2:dec_activity}
  \end{method}
\end{figure}

Method \cleanup{i} is called by either \CL{i} as just explained, when $C[i]$ is decremented to zero, or by \Swap{i} (\autoref{line:swap:confirm_quiescence}) when reading zero from $C[i]$.
In both cases, the calling process $p$ has (at some point before invoking \cleanup{i}) the responsibility to reset $L[i].\Tag$ to zero, unless $L[i]$ changes again.

\begin{figure}
  \begin{method}{\W{cleanup($i \in \set{1, \dots, m}$)}}
    \HI{
      $(val, tag) \gets L[i].\Read{}$\;\label{line:cleanup:first_read}
      \lIf{$tag = 0$}{\Return}\label{line:cleanup:zero_tag_return}
      $R[tag].\FAI{}$\;\label{line:cleanup:inc_refcount} %
      \If{$L[i].\Read{} = (val, tag)$}{\label{line:cleanup:second_read}
        \If{$C[i].\Read{} = 0$}{\label{line:cleanup:confirm_quiescence}
          \lHI{}\WHI{\If{$\HI{L[i].\CAS{(val, tag), (val, 0)}} = \True$\label{line:cleanup:cas}}{
            $R[tag].\FAD{}$\label{line:cleanup:dec_prev}
          }}
        }
      }
      $R[tag].\FAD{}$\;\label{line:cleanup:final_return}
    }
  \end{method}
\end{figure}

Lines~\ref{line:cleanup:first_read},~\ref{line:cleanup:inc_refcount}, and~\ref{line:cleanup:second_read} work the same way as lines~\ref{line:ll:first_read},~\ref{line:ll:inc_refcount}, and~\ref{line:ll:second_read} of \LL{} to try to protect the current tag $g$ stored in $L[i].\Tag$, preventing $p$ from resetting $L[i].\Tag$ after an ABA\@.
If $g$ is zero then $p$ can return immediately (\autoref{line:cleanup:zero_tag_return}).
If $L[i].\Tag$ changes from $g$, then \CL{i} no longer needs to reset $L[i].\Tag$, so it simply decrements its usage count, ending its protection of tag $g$ (\autoref{line:cleanup:final_return}).
Otherwise, $p$ next double-checks that the activity counter $C[i]$ is still zero (\autoref{line:cleanup:confirm_quiescence}).
This second check of $C[i]$ ensures that any process which changes $L[i].\Tag$ does so after $p$'s execution of \autoref{line:cleanup:confirm_quiescence}, while $g$ is protected.
Lines~\ref{line:cleanup:cas} and~\ref{line:cleanup:dec_prev} work similarly to lines~\ref{line:sc2:first_cas} and~\ref{line:sc2:dec_prev} of \SC{}: $p$ attempts to \CAS{} $L[i].\Tag$ from $g$ to zero, and if successful $p$ then decrements tag $g$'s usage count since $g$ is no longer used in $L[i].\Tag$.
(Note that, in the case of a successfully reset tag, $p$ decrements $R[g]$ twice, in lines~\ref{line:cleanup:dec_prev} and~\ref{line:cleanup:final_return}:
The first decrements $L_i$'s usage count associated with the initial lock in \autoref{line:swap:lock} or \autoref{line:sc2:lock}, whereas the second ends $p$'s protection of the tag.)

\subsubsection{Read() and VL() Methods}

The \Read{} and \VL{} methods are the same as in~\cite{BHW2026a}.

A \Read{i} call simply returns the contents of $L[i].\val$, as it has no need to do anything with the \Tag field.

\begin{figure}
  \begin{method}{Read($i \in \set{1, \dots, m}$)}
    \Return{$L[i].\Read{}.\val$}\label{line:read}
  \end{method}
\end{figure}

Conversely, a \VL{i} call only checks $L[i].\Tag$ in \autoref{line:vl:read}, linearizing at the point of the read.
If \VL{i} is called by a process $p$ with an outstanding \LL{i} (which is the intended case, as it would otherwise return in \autoref{line:vl:early_return}), then $p$ would have already protected the tag that it read earlier and stored in $d[i].\Tag$.
This protection prevents ABAs while $p$ has an outstanding \LL{i}, and therefore if $L[i].\Tag = d[i].\Tag$ then it also follows that $L[i]$ did not change in between the linearization points of \LL{i} and \VL{i}.
\VL{i} also checks if $L[i].\Tag = 0$ because of the same edge case discussed for \SC{} in \autoref{sec:low-level:sc}.
If $L[i].\Tag$ is a different nonzero value, then $p$ does not have an open link to $L_i$, and so $p$ calls \CL{i} in \autoref{line:vl:clear} to also clear its outstanding \LL{i}.

\begin{figure}
  \begin{method}{VL($i \in \set{1, \dots, m}$)}
    \lIf{$d[i] = \bot$}{\Return{\False}}\label{line:vl:early_return}
    \lIf{$L[i].\Read{}.\Tag \in \set{d[i].\Tag\HI{, 0}}$}{\Return{\True}}\label{line:vl:read}\label{line:vl:return_true}
    \CL{i}\;\label{line:vl:clear}
    \Return{\False}\label{line:vl:return_false}
  \end{method}
\end{figure}

\renewcommand{\theHdefinition}{\thesection.\thetheorem}
\renewcommand{\theHtheorem}{\thesection.\thetheorem}
\renewcommand{\theHlemma}{\thesection.\thetheorem}
\renewcommand{\theHcorollary}{\thesection.\thetheorem}

\section{Analysis}\label{sec:analysis}
This section formally proves \autoref{thm:write:main}.
Several proofs closely follow the format of those of~\cite{BHW2026a-full} (the full version of~\cite{BHW2026a}), but with significant changes throughout due to adding the \Swap{} operation and making the required changes to \SC{} and \CL{}.

\setcounter{theorem}{2}

\begin{lemma}\label{lem:write:cl_clears_di}
  When a process responds from a \CL{i}, \Swap{i, \cdot}, or \SC{i, \cdot} call, its value of $d[i]$ is $\bot$.
\end{lemma}
This lemma corresponds to Lemma~1 of~\cite{BHW2026a-full}, and its proof is mostly unchanged, except that it also applies to \Swap{i, \cdot}.
\begin{proof}
  For \CL{i}, if the calling process $p$ returns in \autoref{line:cl2:early_return}, then the lemma is satisfied by the if-condition in that line.
  Otherwise, $p$ sets $d[i]$ to $\bot$ in \autoref{line:cl2:clear_di} and does not change it again (including in \cleanup{i}, if called) until its response.

  For \Swap{i, \cdot}, if $d[i] \neq \bot$ upon invocation, then the calling process $p$ calls \CL{i} in \autoref{line:swap:clear}, which by the above ensures that $d[i] = \bot$ by its response.
  The rest of the call does not change $d[i]$ again, including in a possible call to \cleanup{i}.

  For \SC{i, \cdot}, if the calling process $p$ returns in \autoref{line:sc2:early_return}, then the lemma is satisfied by the if-condition in that line.
  Otherwise, $p$ calls \CL{i} in \autoref{line:sc2:clear}, which by the above ensures that $d[i] = \bot$ by its response, and then immediately returns.
\end{proof}

\begin{lemma}\label{lem:write:nonbot_pending_link}
  If $d[i] \neq \bot$ for a process $p$ while $p$ does not have a pending method call on $L_i$, then $p$ has an outstanding \LL{i}.
\end{lemma}
This lemma corresponds to Lemma~2 of~\cite{BHW2026a-full}, and its proof is mostly unchanged.
\begin{proof}
  Only \LL{i} can set $d[i]$ to a non-$\bot$ value.
  By \autoref{lem:write:cl_clears_di}, the \CL{i}, \Swap{i, \cdot}, and \SC{i, \cdot} methods ensure $d[i] = \bot$ by their respective responses.
  If a \VL{i} call returns \False, then either $d[i] = \bot$ in \autoref{line:vl:early_return}, or it calls \CL{i} in \autoref{line:vl:clear}, so in both cases $d[i] = \bot$ by the point it returns \False.
  Thus, if $d[i] \neq \bot$ while $p$ is idle, then $p$ has called \LL{i} and not subsequently called \Swap{i, \cdot}, \SC{i, \cdot}, \CL{i}, or a \VL{i} call which returns \False, so by \autoref{def:write:outstanding_ll} $p$ has an outstanding \LL{i}.
\end{proof}

\subsection{Contributions to Counters}\label{sec:analysis:contributions}
The following definition corresponds to Definition~3 of~\cite{BHW2026a-full}, but is more complicated because LL/SC objects can now contribute to elements of $R$, with said contribution able to transfer between processes and LL/SC objects.
Intuitively, a contribution by $p$ to $R[g]$ represents $p$'s responsibility for eventually decrementing the value of $R[g]$.
That responsibility is initially generated by $p$'s increment of $R[g]$ (from a \FAI{} or a successful $CAS(0, 1)$).
When a process installs a new tag to $L[i].\Tag$ for some object $L_i$, it transfers that responsibility to $L_i$.
When $L[i].\Tag$ is subsequently replaced by a process, the responsibility transfers to the replacing process, which then performs the decrement of $R[g]$.
\begin{definition}[Contribution]\label{def:write:contribute}
  Given a point $t$, a process $p$, and element $C[i]$, $p$ \emph{contributes $j$} to $C[i]$'s value, if $j$ is the number of \FAI{} minus \FAD{} operations that $p$ has executed on $C[i]$ before $t$.

  For each element $R[g]$, processes increase their contributions by performing a \FAI{} or a successful \CAS{0, 1} operation on $R[g]$, and decrease their contribution to $R[g]$ by performing a \FAD{} operation on $R[g]$.
  In addition, contributions to $R[g]$ can also transfer between processes and LL/SC objects, without changing the total contribution to or the value of $R[g]$, as follows:
  \begin{itemize}
    \item Suppose a process $p$ executes an $L[i].\Swap{(\cdot, g)}$ operation in \autoref{line:swap:swap} or a successful $L[i].\CAS{\cdot, (\cdot, g)}$ operation in \autoref{line:sc2:first_cas} or~\ref{line:sc2:second_cas} at point $t$.
    Then, $p$'s contribution to $R[g]$'s value decrements by one and $L_i$'s contribution to $R[g]$'s value increments by one at point $t$.
    \item Suppose a process $p$ reads $(\cdot, g)$ from an $L[i].\Swap{\cdot}$ operation in \autoref{line:swap:swap}, or executes a successful $L[i].\CAS{(\cdot, g), \cdot}$ operation in \autoref{line:sc2:first_cas} or~\ref{line:cleanup:cas} at point $t$.
    Then, $p$'s contribution to $R[g]$'s value increments by one and $L_i$'s contribution to $R[g]$'s value decrements by one at point $t$.
  \end{itemize}
\end{definition}

\begin{lemma}\label{lem:write:process_contribution}
  When a process $p$ invokes or responds from a method call, including the helper method \cleanup{},
\begin{enumerate}[(a)]
    \item if $d[i] = \bot$, then $p$ contributes zero to $C[i]$, and otherwise $p$ contributes one to $C[i]$, and\label{item:write:idle_contribution:C}
    \item the total contribution to $R[g]$ by $p$, for $g \in \set{1, \dots, 4(k + 1)n}$, equals the number of non-$\bot$ values in $p$'s dictionary $d$ which have $g$ as their tag.\label{item:write:idle_contribution:R}
  \end{enumerate}
  At all points in the execution,
  \begin{enumerate}[(a)]
    \setcounter{enumi}{2}
    \item the number of objects $L_i$ for which $d[i] \neq \bot$ for $p$ is at most $k$,\label{item:write:d_size}
    \item $p$ contributes either zero or one to each individual element of $C$, and\label{item:write:contribution:C}
    \item $p$ contributes a total value in $\set{0, \dots, k + 2}$ to the elements of $R$, with a non-negative contribution to each individual element.\label{item:write:contribution:R}
  \end{enumerate}
\end{lemma}
This lemma corresponds to Lemma~4 of~\cite{BHW2026a-full}, but~(\ref{item:write:contribution:R}) is a per-process bound and includes a lower bound of zero, with a global lower bound proven later in \autoref{cor:write:non-negative_contribution}.
\begin{proof}
  Note that by \autoref{def:write:contribute}, $p$'s contributions to $R$ and $C$, as well as its dictionary $d$, do not change while $p$ is idle.
  Thus, if~(\ref{item:write:idle_contribution:C}) through~(\ref{item:write:contribution:R}) hold when $p$ responds from a method call and becomes idle, then they also hold until the point when $p$ next invokes a method.

  As such, we prove~(\ref{item:write:idle_contribution:C}) and~(\ref{item:write:idle_contribution:R}) for a process $p$ by induction over its method calls, including the helper method \cleanup{}, while proving that~(\ref{item:write:d_size}) through~(\ref{item:write:contribution:R}) hold at all points within each of $p$'s method calls.

  Initially, $p$ contributes zero to elements of $C$ and $R$ and its dictionary $d$ is empty (all values are $\bot$), so~(\ref{item:write:idle_contribution:C}) through~(\ref{item:write:contribution:R}) hold before $p$'s first method call.

  Suppose that~(\ref{item:write:idle_contribution:C}) through~(\ref{item:write:contribution:R}) hold for $p$ before the invocation of a given method: we will prove that~(\ref{item:write:d_size}) through~(\ref{item:write:contribution:R}) hold throughout the method call, and that~(\ref{item:write:idle_contribution:C}) and~(\ref{item:write:idle_contribution:R}) hold at the call's response.

  If the method call is \Read{i}, then~(\ref{item:write:idle_contribution:C}) through~(\ref{item:write:contribution:R}) hold because \Read{} does not affect $p$'s contributions or its dictionary $d$.

  We next consider the \CL{i} method.
  If the call returns immediately in \autoref{line:cl2:early_return}, then neither $p$'s contribution nor $d$ change, so~(\ref{item:write:idle_contribution:C}) through~(\ref{item:write:contribution:R}) remain true.
  Otherwise, $d[i] \neq \bot$ at invocation, so by~(\ref{item:write:idle_contribution:C}), $p$'s contribution to $C[i]$ is one at invocation.
  Since the \cleanup{i} method is called as the last step, if it is called at all, it suffices to show that~(\ref{item:write:idle_contribution:C}) and~(\ref{item:write:idle_contribution:R}) hold immediately after the \FAD{} on $C[i]$ in \autoref{line:cl2:dec_activity}, and~(\ref{item:write:d_size}) through~(\ref{item:write:contribution:R}) hold until just after that decrement.
  Since $p$ does not set any non-$\bot$ value in $d$ throughout the method call,~\textbf{(\ref{item:write:d_size}) remains satisfied}.
  Since $p$ resets $d[i]$ to $\bot$ in \autoref{line:cl2:clear_di} and decrements $C[i]$ in \autoref{line:cl2:dec_activity}, $p$'s contribution to $C[i]$ becomes zero which \textbf{satisfies~(\ref{item:write:idle_contribution:C})} after the decrement.
  In particular, $p$'s contribution is either zero or one throughout the call, so~\textbf{(\ref{item:write:contribution:C}) is satisfied} throughout the call.

  If $d[i].\Tag = 0$ at invocation, then $p$ does not make any change to $R$ in \autoref{line:cl2:dec_refcount}, so~(\ref{item:write:idle_contribution:R}) and~(\ref{item:write:contribution:R}) remain true throughout the call.
  Otherwise, $d[i].\Tag$ is nonzero at invocation, so by~(\ref{item:write:idle_contribution:R}) $p$ contributes at least one to $R[d[i].\Tag]$ at invocation.
  In this case, $p$ decrements $R[d[i].\Tag]$ in \autoref{line:cl2:dec_refcount}, satisfying~(\ref{item:write:idle_contribution:R}) again after \autoref{line:cl2:clear_di}.
  In particular, since $p$'s total contribution to $R$ is non-negative at invocation by~(\ref{item:write:contribution:R}), its contribution to $R[d[i].\Tag]$ is positive at invocation by~(\ref{item:write:idle_contribution:R}), and it only decrements $R[d[i].\Tag]$,~(\ref{item:write:contribution:R}) is satisfied throughout the method call.
  Thus, both cases \textbf{satisfy~(\ref{item:write:idle_contribution:R}) and~(\ref{item:write:contribution:R})}.

  Next, consider the \cleanup{i} helper method.
  Since it does not modify $d$ or $C$, \textbf{(\ref{item:write:idle_contribution:C}),~(\ref{item:write:d_size}), and~(\ref{item:write:contribution:C}) remain satisfied}.
  If $p$ responds in \autoref{line:cleanup:final_return}, then it increments and later decrements the value of $R[tag]$ for its local variable $tag$, in lines~\ref{line:cleanup:inc_refcount} and~\ref{line:cleanup:final_return}, which exactly counteract each other's effect.
  If $p$ executes a successful \CAS{} in \autoref{line:cleanup:cas}, then (by \autoref{def:write:contribute}) it increments its contribution to $R[tag]$ for its local variable $tag$, and then decrements that contribution to $R[tag]$ when $p$ executes its next step in \autoref{line:cleanup:dec_prev}.
  Since the method does not modify $d$, and any contribution increments are followed by contribution decrements of the same element of $R$,~\textbf{(\ref{item:write:idle_contribution:R}) holds} at $p$'s response in \autoref{line:cleanup:final_return} or~\ref{line:cleanup:zero_tag_return}.
  From~(\ref{item:write:contribution:R}) and~(\ref{item:write:d_size}), $p$'s total contribution to elements of $R$ is at most $k$ when invoking \cleanup{i}.
  Since $p$'s contribution increases at most twice in \cleanup{i}, and since all decrements of contributions to elements of $R$ follow increments of that same element in the same method,~\textbf{(\ref{item:write:contribution:R}) remains satisfied}.

  Since the \VL{} method does not change $p$'s contribution to $C$ or $R$ or its dictionary $d$ except via its internal \CL{} call,~(\ref{item:write:idle_contribution:C}) through~(\ref{item:write:contribution:R}) hold for \VL{} by the above arguments for \CL{} and \cleanup{}.

  We next consider the \Swap{i, \cdot} method, before the possible call to \cleanup{i} in \autoref{line:swap:confirm_quiescence}.
  We have that \textbf{(\ref{item:write:idle_contribution:C}),~(\ref{item:write:d_size}), and~(\ref{item:write:contribution:C}) remain true} throughout the method call, since $p$'d dictionary $d$ and its contribution to $C$ do not change in the method.
  By~(\ref{item:write:idle_contribution:R}) and~(\ref{item:write:d_size}) from the inductive hypothesis, $p$ contributes at most $k$ to elements of $R$ at invocation.
  Then, $p$'s contribution to $R[tag']$ increments by one at a successful \CAS{} in \autoref{line:swap:lock}, and decrements by one in \autoref{line:swap:swap}.
  If the value $tag$ read from $L[i].\Tag$ in \autoref{line:swap:swap} is nonzero, then $p$'s contribution to $R[tag]$ increments by one in \autoref{line:swap:swap} and decrements by one in \autoref{line:swap:dec_prev}.
  In both cases, $p$'s contributions to elements of $R$ before \autoref{line:swap:confirm_quiescence} changes back to what they were at invocation, \textbf{satisfying~(\ref{item:write:idle_contribution:R})} before \autoref{line:swap:confirm_quiescence}.
  Since $p$'s total contribution to $R$ does not increase by more than one relative to invocation throughout the method call, and all increments are then cancelled out for the same element of $R$,\textbf{~(\ref{item:write:contribution:R}) remains satisfied} throughout the method call.

  The arguments for the \SC{i, \cdot} method are similar to those for \Swap{i, \cdot}, but there are more cases to consider.
  Here, we need to consider steps, up to just before it calls \CL{i} in \autoref{line:sc2:clear}.
  Since the steps of \SC{} excluding \CL{} do not change the values in $d$ or $p$'s contributions to $C$, it again follows that \textbf{(\ref{item:write:idle_contribution:C}),~(\ref{item:write:d_size}), and~(\ref{item:write:contribution:C}) remain true} throughout those steps.

  If the \SC{i, \cdot} call returns in \autoref{line:sc2:early_return}, then~(\ref{item:write:idle_contribution:R}) and~(\ref{item:write:contribution:R}) also hold throughout the call since the call does not change $d$ or $R$.
  Otherwise, $d[i] \neq \bot$ at invocation, so we consider the lines up until \autoref{line:sc2:clear}.
  Again by~(\ref{item:write:idle_contribution:R}) and~(\ref{item:write:d_size}) from the inductive hypothesis, $p$ contributes a non-negative integer of at most $k$ to the elements of $R$ at \SC{i, \cdot}'s invocation.
  Thus, when $p$ breaks out of its while-loop via a successful \CAS{0, 1} in \autoref{line:sc2:lock}, its contribution to $R[tag']$ increments by one, and its total contribution to $R$ is at most $k+1$ immediately after that line.
  By \autoref{def:write:contribute}, that contribution lasts until either a successful \CAS{} in \autoref{line:sc2:first_cas} or~\ref{line:sc2:second_cas}, or upon a \FAD{} in \autoref{line:sc2:unlock}.

  If $p$ has a successful \CAS{} in \autoref{line:sc2:first_cas}, and if $d[i].\Tag \neq 0$, then $p$'s contribution to $R[tag']$ is decremented and its contribution to $R[d[i].\Tag]$ is incremented at that point, which does not change $p$'s total contribution to $R$.
  Exactly in that case, $p$ executes the $R[d[i].\Tag].\FAD{}$ in \autoref{line:sc2:dec_prev}, undoing the effect of incrementing its contribution to $R[d[i].\Tag]$.
  By inspection of the code and \autoref{def:write:contribute}, $p$'s contribution to $R[tag']$ decrements by one in exactly either \autoref{line:sc2:first_cas},~\ref{line:sc2:second_cas}, or~\ref{line:sc2:unlock}, bringing $p$'s total contribution to $R$ back to at most $k$.

  Since all increments of elements of $R$ by $p$ are then decremented for the same element in the method, its contributions immediately before \autoref{line:sc2:clear} are the same as at invocation, thus \textbf{satisfying~(\ref{item:write:idle_contribution:R})} before \autoref{line:sc2:clear}.
  In particular, $p$'s contribution to $R[tag']$ increments by one and then decrements by one, with the decrement possibly happening simultaneously with an increment of the contribution to $R[d[i].\Tag]$ followed by a corresponding decrement.
  Therefore, because $p$'s total contributions to $R$ at invocation are in $\set{0,\dots,k}$, and from~(\ref{item:write:contribution:R}) $p$ contributes a non-negative value to each element of $R$ at invocation,~\textbf{(\ref{item:write:contribution:R}) remains satisfied} throughout the call.

  It remains to show~(\ref{item:write:idle_contribution:C}) through~(\ref{item:write:contribution:R}) for an \LL{i} call by $p$.
  Since they each remain true after \CL{} calls by the above arguments, they each remain true by the time the \LL{i} call reaches \autoref{line:ll:inc_activity}.

  Since, by \autoref{lem:write:cl_clears_di}, the \CL{i} call in \autoref{line:ll:reset_link} ensures that $d[i] = \bot$, it follows by~(\ref{item:write:idle_contribution:C}) that $p$'s contribution to $C[i]$ immediately before \autoref{line:ll:inc_activity} is zero, and hence its contribution immediately after the \FAI{} is one.
  In particular, at all points $p$'s contribution to $C[i]$ is either zero or one, \textbf{satisfying~(\ref{item:write:contribution:C})} throughout the call.
  After executing \autoref{line:ll:set_di},~(\ref{item:write:idle_contribution:C}) becomes true again, and remains true until $p$'s possible invocation of \CL{i} in \autoref{line:ll:clear}, then remains true at its response by the above arguments for \CL{}, and therefore~\textbf{(\ref{item:write:idle_contribution:C}) is true} at any possible response from \LL{i}.
  (Note that \autoref{line:ll:reset_tag}, if executed, only changes the value of $d[i]$ from one non-$\bot$ value to another.)
  Recall from \autoref{sec:write:pending_links} that $p$ is required to have at most $k - 1$ outstanding \LL{} operations to objects other than $L_i$ before invoking \LL{i}.
  Thus, by \autoref{lem:write:nonbot_pending_link}, there are at most $k - 1$ objects $L_j \neq L_i$ such that $d[j] \neq \bot$ for $p$ before the call's invocation.
  Since $d[i] = \bot$ after \autoref{line:ll:reset_link}, it follows that after executing \autoref{line:ll:set_di}, there are at most $k$ non-$\bot$ values in $d$, and so~\textbf{(\ref{item:write:d_size}) remains true} throughout the call.

  Since~(\ref{item:write:idle_contribution:R}) holds immediately before $p$ executes \autoref{line:ll:inc_activity}, and the number of non-$\bot$ values in $d$ is less than $k$ at that point, it follows that \autoref{line:ll:inc_refcount} does not increase $p$'s contribution to $R$ past $k$.
  By the inductive hypothesis and that \CL{} preserves~(\ref{item:write:contribution:R}),~(\ref{item:write:contribution:R}) holds from invocation until just before \autoref{line:ll:inc_refcount} (or a response in \autoref{line:ll:first_zero_tag}).
  It follows that $p$'s contribution to $R[tag]$ is at least $1$ after the \FAI{} in \autoref{line:ll:inc_refcount}, and therefore remains non-negative after a possible \FAD{} in \autoref{line:ll:dec_refcount}.
  Therefore, $p$'s contribution to $R$ is in $\set{0, \dots, k}$ throughout the \LL{i} call, with a non-negative contribution to each element which is more than enough to \textbf{satisfy~(\ref{item:write:contribution:R})} throughout the call.
  Finally, we can verify that~\textbf{(\ref{item:write:idle_contribution:R}) holds} at each relevant response point and the possible invocation of \CL{i} in \autoref{line:ll:clear}, using that it holds immediately after \autoref{line:ll:reset_link} with $d[i] = \bot$ at that point.
  In \autoref{line:ll:first_zero_tag}, $d[i].\Tag = 0$ and $R$ is untouched by $p$; in lines~\ref{line:ll:matching_tag} and~\ref{line:ll:clear}, $d[i].\Tag \neq 0$ and $R[tag] = R[d[i].\Tag]$ is incremented; and in \autoref{line:ll:second_zero_tag}, $d[i].\Tag = 0$ and the increment of $R[tag]$ is reversed.
\end{proof}

\begin{lemma}\label{lem:write:C_value_equals_contributions}
  The value of each element in $C$ equals the sum of the contributions from each process (i.e., no element overflows).
\end{lemma}
This lemma corresponds to part of Lemma~6 of~\cite{BHW2026a-full}, with the part referring to $R$ proven later in \autoref{lem:write:R_value_and_mutex_lock}(\ref{item:write:R_value_equals_contributions}).
\begin{proof}
  From \autoref{def:write:contribute}, each \FAI{} to an element of $C$ corresponds to an increment of the total contribution to that element.
  Similarly, each \FAD{} to an element of $C$ corresponds to a decrement of the total contribution to that element.
  Since no other operation affects the value of elements of $C$, it suffices to show that the total contribution to each element remains in the range $[0, n]$, thus never underflowing or overflowing the element's value.
  From \autoref{lem:write:process_contribution}(\ref{item:write:contribution:C}), process contributions to $C$ are either zero or one at all times.
  Therefore, each process contributes at most one to the value of each element of $C$, and since objects do not contribute to elements of $C$, it follows that the maximum contribution to any given element of $C$ is $n$.
  It follows that the total contribution each element of $C$ is always in the range $[0, n]$, implying the lemma.
\end{proof}

\begin{lemma}\label{lem:write:contribute_while_in_L}
  At point $t$, $L[i].\Tag = g \neq 0$ if and only if $L_i$ contributes one to the value of $R[g]$.
  For all $g' \notin \set{L[i].\Tag, 0}$, $L_i$ contributes zero to the value of $R[g']$ at point $t$.
\end{lemma}
This is a new result, which we use to ensure correctness of the \Swap{} method.
\begin{proof}
  The only lines that can change $L_i$'s contribution to the value of an element of $R$, by \autoref{def:write:contribute}, are lines~\ref{line:swap:swap},~\ref{line:sc2:first_cas},~\ref{line:sc2:second_cas}, and~\ref{line:cleanup:cas}.
  These are also the only lines that can change the value of $L[i]$.
  From \autoref{def:write:contribute} and inspection of the code, the cases in which an object increments its contribution to $R[g]$ for nonzero tag $g$ and the cases in which $L[i].\Tag$ changes to nonzero tag $g$ are the same as each other:
  Both occur exactly in the event of an $L[i]$.\Swap{(\cdot, g)} operation in \autoref{line:swap:swap} or a successful $L[i]$.\CAS{\cdot, (\cdot, g)} operation in \autoref{line:sc2:first_cas} or~\ref{line:sc2:second_cas}.
  Similarly, from \autoref{def:write:contribute} and inspection of the code, the cases in which an object \emph{decrements} its contribution to $R[g]$ for nonzero tag $g$ and the cases in which $L[i].\Tag$ changes \emph{from} nonzero tag $g$ are the same as each other:
  Both occur exactly in the event of an $L[i].\Swap{\cdot}$ operation in \autoref{line:swap:swap} that returns $(\cdot, g)$, or a successful $L[i].\CAS{(\cdot, g), \cdot}$ operation in \autoref{line:sc2:first_cas} or~\ref{line:cleanup:cas}.
  The above are not (yet) mutually exclusive: if a \Swap{} or successful \CAS{} on $L[i]$ `changes' $L[i].\Tag$ from $g$ to $g$, then by \autoref{def:write:contribute} $L_i$'s contribution to tag $g$ both increments and decrements by one, and therefore does not change.
  (We later show in \autoref{lem:write:write_sc_changes_tag} that this cannot occur, but that result depends on this lemma.)

  Initially, $L[i].\Tag = 0$ for all $i$, and no object contributes anything to any element of $R$.
  From the above observations, it follows that $L_i$ increments its contribution to $R[g]$'s value for nonzero $g$ exactly when $L[i].\Tag$ changes to $g$ from a different tag.
  Similarly, $L_i$ decrements its contribution to $R[g]$'s value exactly when $L[i].\Tag$ changes from $g$ to different tag.
  The first part of the lemma thus follows by induction.
  Since $L[i].\Tag$ only has one value, $L_i$ can therefore only contribute to at most one element of $R$ at a time, namely the element $R[g]$ such that $L[i].\Tag = g \neq 0$, proving the second part.
\end{proof}

\begin{corollary}\label{cor:write:non-negative_contribution}
  At all points of the execution, the total contribution to any given element of $C$ or $R$ is non-negative, and it is positive if any process or object contributes a positive value to that element.
\end{corollary}
This statement is somewhat analogous to Lemma~5 of~\cite{BHW2026a-full}, but proven differently.
\begin{proof}
  Since only processes contribute to $C$, and by \autoref{lem:write:process_contribution}(\ref{item:write:contribution:C}) they each contribute either zero or one to each element, it follows that the total contribution to $C$ is non-negative, and positive if any process contributes one.
  The statement for $R$ follows from the combination of \autoref{lem:write:process_contribution}(\ref{item:write:contribution:R}) and \autoref{lem:write:contribute_while_in_L}.
\end{proof}

\subsection{Locking Tags}

An important new concept in this swappable LL/SC implementation is that of \emph{locking} tags.
The \SC{} and \Swap{} methods attempt to lock a tag as defined below to ensure that they do not choose the same new tag as a concurrent \SC{} or \Swap{} call, which could lead to ABAs as argued in \autoref{sec:changes}.
\begin{definition}[Locked tag]\label{def:write:lock}
  A process $p$ \emph{locks} tag $g \in \set{1, \dots, 4(k + 1)n}$ for $L_i$ when $p$ has a successful \CAS{0, 1} on $R[g]$ in either \autoref{line:swap:lock} or~\ref{line:sc2:lock}, as part of an \SC{i, \cdot} or \Swap{i, \cdot} call.
  The tag then becomes \emph{unlocked} when the total contribution to it according to \autoref{def:write:contribute} becomes zero.
\end{definition}

\begin{lemma}\label{lem:write:locked_while_in_L}
  While $L[i].\Tag = g \neq 0$, tag $g$ is locked for $L_i$.
  When a \Swap{} or a successful \CAS{} next takes effect on $L[i]$, tag $g$ remains locked until the process that executes that \Swap{} or \CAS{} takes its next step.
\end{lemma}
\begin{proof}
  Initially, all \Tag fields are set to zero.
  Thus, if $L[i].\Tag = g \neq 0$ at point $t$, then there is a point $t' < t$ at which a \Swap{} in \autoref{line:swap:swap} or a successful \CAS{} in either \autoref{line:sc2:first_cas} or~\ref{line:sc2:second_cas} is executed by some process $p$, changing $L[i].\Tag$ to $g$.
  Let $t'$ be the latest such point, so that $L[i].\Tag = g$ throughout $(t', t]$.
  There is a point $t_L < t'$ at which $p$ executes a successful \CAS{} in \autoref{line:swap:lock} or~\ref{line:sc2:lock} on $R[g]$, changing it from zero to one, and by \autoref{def:write:contribute} and \autoref{lem:write:process_contribution}(\ref{item:write:contribution:R}) increasing $p$'s contribution to $R[g]$ from zero to one.
  By \autoref{def:write:lock}, $p$ locks tag $g$ at point $t_L$.
  From \autoref{def:write:contribute}, $p$'s contribution to $R[g]$ does not change from $t_L$ to $t'$, $p$'s contribution is transferred to $L_i$ at point $t'$, and $L_i$'s contribution does not change from $t'$ to $t$.
  From \autoref{cor:write:non-negative_contribution}, there is a non-negative contribution to $R[g]$ throughout $(t_L, t]$.
  Therefore, from \autoref{def:write:lock}, tag $g$ remains locked throughout $(t_L, t]$.

  Now suppose that point $t^\star$ is the first point after $t$ at which a \Swap{} in \autoref{line:swap:swap} or a successful \CAS{} in either \autoref{line:sc2:first_cas} or~\ref{line:cleanup:cas} takes effect on $L[i]$, and let $q$ be the process performing that step.
  $L[i].\Tag = g$ throughout $(t', t^\star)$, $L_i$'s contribution transfers to $q$ at point $t^\star$ (from \autoref{def:write:contribute}), and this transferred contribution remains until $q$ its next step, which is \autoref{line:swap:dec_prev},~\ref{line:sc2:dec_prev}, or~\ref{line:cleanup:dec_prev}, at point $t_U > t^\star$.
  From \autoref{cor:write:non-negative_contribution}, there is a positive contribution to $R[g]$ throughout $(t_L, t_U)$.
  The final part of the lemma then follows from \autoref{def:write:lock}.
\end{proof}

\begin{lemma}\label{lem:write:R_value_and_mutex_lock}
  At each point in the execution,
  \begin{enumerate}[(a)]
    \item the value of each element in $R$ equals the sum of the contributions from each process and LL/SC object,\label{item:write:R_value_equals_contributions}
    \item while a tag is locked, no process can lock it a second time, and\label{item:write:mutex_lock}
    \item each process contributes at most one to each element of $R$, unless a contribution was transferred from an object to the process in either \autoref{line:swap:swap},~\ref{line:sc2:first_cas}, or~\ref{line:cleanup:cas}.
    In the latter case, the process can contribute at most two to that element until it next executes \autoref{line:swap:dec_prev},~\ref{line:sc2:dec_prev}, or~\ref{line:cleanup:dec_prev}, respectively.\label{item:write:one_contribution_per_tag}
  \end{enumerate}
\end{lemma}
Statement~(\ref{item:write:R_value_equals_contributions}) corresponds to part of Lemma~6 from~\cite{BHW2026a-full}, with the part referring to $C$ proven earlier in \autoref{lem:write:C_value_equals_contributions}.
\begin{proof}
  We prove the three parts as a combined invariant by induction over the execution.
  They each hold at the beginning of the execution, since no process or object contributes to $R$ and no tag is locked initially.

  We first show~(\ref{item:write:mutex_lock}), by showing that if a tag $g$ becomes locked at point $t$, and~(\ref{item:write:R_value_equals_contributions}) holds at all points prior to $t$, then $g$ is not locked immediately before $t$.
  From \autoref{def:write:lock}, there is a nonzero total contribution to $R[g]$ throughout the interval in which tag $g$ is locked.
  Since point $t$ is the point of a successful \CAS{0, 1} on $R[g]$, it follows that $R[g] = 0$ immediately before point $t$.
  From~(\ref{item:write:R_value_equals_contributions}), it follows that no process or object contributes to $R[g]$ immediately before $t$, and therefore that $g$ is not locked immediately before $t$.

  We next show that~(\ref{item:write:one_contribution_per_tag}) holds at point $t$, given that~(\ref{item:write:R_value_equals_contributions}),~(\ref{item:write:mutex_lock}), and~(\ref{item:write:one_contribution_per_tag}) hold at all points prior to point $t$.
  Suppose that a process $p$ increments its contribution to $R[g]$ at point $t$ (otherwise~(\ref{item:write:one_contribution_per_tag}) holds immediately by the inductive hypothesis).
  If this comes from a transfer from an object, i.e., in either \autoref{line:swap:swap},~\ref{line:sc2:first_cas}, or~\ref{line:cleanup:cas}, then by the inductive hypothesis for~(\ref{item:write:one_contribution_per_tag}), $p$ contributes at most one per tag immediately before point $t$.
  Therefore, $p$ contributes at most two per tag immediately after point $t$.
  In all transferred cases, $p$ decrements the transferred tag in the next line (\autoref{line:swap:dec_prev},~\ref{line:sc2:dec_prev}, or~\ref{line:cleanup:dec_prev}, respectively), from which~(\ref{item:write:one_contribution_per_tag}) follows.

  Thus, we next consider the cases in which $p$ directly increments $R[g]$ at point $t$, in either \autoref{line:ll:inc_refcount},~\ref{line:swap:lock},~\ref{line:sc2:lock}, or~\ref{line:cleanup:inc_refcount}: in all cases, we will show that $p$ contributes exactly one to $R[g]$ immediately after point $t$.
  For lines~\ref{line:swap:lock} and~\ref{line:sc2:lock}, since they are \CAS{0, 1} operations,~(\ref{item:write:one_contribution_per_tag}) follows from the inductive hypothesis that~(\ref{item:write:R_value_equals_contributions}) holds immediately before $t$, and that a total contribution of zero to $R$ implies zero contribution from each process (from \autoref{cor:write:non-negative_contribution}).

  The remaining (and most complicated) cases are that $p$ executes \autoref{line:ll:inc_refcount} of an \LL{i} call or \autoref{line:cleanup:inc_refcount} of a \cleanup{i} call at point $t$, which we consider together.
  Suppose for the purpose of proving a contradiction that $p$ already contributes at least one to $R[g]$ immediately before point $t$.
  It follows that there is a point $t' < t$ at which $p$ incremented its contribution to $R[g]$, without decrementing its contribution to $R[g]$ in $(t', t)$.

  Observe that the increments of contribution which can happen in lines~\ref{line:swap:lock},~\ref{line:swap:swap},~\ref{line:sc2:first_cas}, and~\ref{line:cleanup:inc_refcount} are necessarily followed by a decrement of contribution in lines~\ref{line:swap:swap},~\ref{line:swap:dec_prev},~\ref{line:sc2:dec_prev}, and~\ref{line:cleanup:final_return}, respectively, in the same method call.
  Similarly, the increment of contribution in \autoref{line:sc2:lock} is necessarily followed by a decrement of contribution in exactly one of either \autoref{line:sc2:first_cas},~\ref{line:sc2:second_cas}, or~\ref{line:sc2:unlock}.
  In all the above-considered cases, $p$ does not execute \autoref{line:ll:inc_refcount} or~\ref{line:cleanup:inc_refcount}.
  Therefore, the only increment of contribution which is not cancelled out before point $t$ is that in \autoref{line:ll:inc_refcount}.
  Moreover, that call must respond in \autoref{line:ll:matching_tag}, otherwise $p$'s contribution is decremented in either \autoref{line:ll:dec_refcount} or \autoref{line:cl2:dec_refcount} from the \CL{} call in \autoref{line:ll:clear}.
  So, there is a point $t' < t$ at which $p$ increments $R[g]$ in \autoref{line:ll:inc_refcount} of an \LL{j} call, and that $p$'s contribution to $R[g]$ stays at one until just before point $t$.

  Since $p$'s \LL{j} call responds in \autoref{line:ll:matching_tag}, there is a point $t^\star \in (t', t)$ at which $p$ reads $g$ from $L[j].\Tag$ in \autoref{line:ll:second_read}.
  From \autoref{lem:write:locked_while_in_L}, $g$ is locked at point $t^\star$ for $L_j$.

  We next establish that $L_j \neq L_i$ by contradiction: suppose instead that $L_i = L_j$.
  Since $p$'s contribution to $R[g]$ is exactly one throughout $(t', t)$, and each method call involves only one element of $d$, \autoref{lem:write:process_contribution}(\ref{item:write:idle_contribution:R}) implies that $d[i].\Tag = g$ at $p$'s method invocations throughout the interval.
  In the case of executing \autoref{line:ll:inc_refcount} at point $t$, $p$ would call \CL{j} in \autoref{line:ll:reset_link} before \autoref{line:ll:inc_refcount}, and decrement $R[g]$ in \autoref{line:cl2:dec_refcount}, contradicting that $p$ does not decrement $R[g]$ throughout $(t', t)$.
  In the case of executing \autoref{line:cleanup:inc_refcount} at point $t$, recall that $p$ calls \cleanup{i} from either \Swap{i, \cdot} or \CL{i}.
  If called from \Swap{i, \cdot}, $p$ previously calls \autoref{line:swap:clear} and calls \CL{i} in that line, since the call occurs after the \LL{i} call which includes point $t'$, and therefore $d[i].\Tag = g$ at method invocations.
  The \CL{i} call then decrements $R[d[i].\Tag] = R[g]$ in line \autoref{line:cl2:clear_di} as $g \neq 0$, which again contradicts that $p$ does not decrement $R[g]$ throughout $(t', t)$.

  At point $t^\dagger \in (t^\star, t)$, $p$ reads $g$ from $L[i].\Tag$ in \autoref{line:ll:first_read}.
  Using \autoref{lem:write:locked_while_in_L} again, it follows that $g$ is locked for $L_i$ at point $t^\dagger$.
  But since $p$ contributes one to $R[g]$ throughout $(t', t) \supset [t^\star, t^\dagger]$, by \autoref{def:write:lock}, $g$ is never unlocked for $L_j$ in that interval.
  Thus, at point $t^\dagger$, $g$ is locked for both $L_i$ and $L_j$, which contradicts~(\ref{item:write:mutex_lock}) before point $t$, and finishes the inductive case for~(\ref{item:write:one_contribution_per_tag}).

  To conclude the proof, we show that~(\ref{item:write:R_value_equals_contributions}) holds at point $t$, given that~(\ref{item:write:mutex_lock}) and~(\ref{item:write:one_contribution_per_tag}) hold \textbf{at} point $t$.
  From \autoref{def:write:contribute}, the total contribution to a given element of $R$ increases or decreases exactly when that element is incremented (including via \CAS{0, 1}) or decremented.
  (In particular, transfers of contribution between processes and objects do not change the total contribution to any element of $R$.)
  Since \autoref{cor:write:non-negative_contribution} establishes that contributions are non-negative, to prove~(\ref{item:write:R_value_equals_contributions}) it suffices to show that the total contribution to a given element of $R$ does not exceed $n + 1$ at point $t$, and thus that the element never overflows.

  From \autoref{lem:write:locked_while_in_L} and~(\ref{item:write:mutex_lock}), a given tag $g$ has a total contribution of at most one from all LL/SC objects at point $t$.
  If at point $t$, no process previously received a transferred contribution to $g$ in either \autoref{line:swap:swap},~\ref{line:sc2:first_cas}, or~\ref{line:cleanup:cas} without having taken its next step of decrementing $R[g]$, then by~(\ref{item:write:one_contribution_per_tag}), each process contributes at most one to $R[g]$.
  Thus, in this case, the total contribution to $R[g]$ is at most $n + 1$, which would imply~(\ref{item:write:R_value_equals_contributions}).

  To finish proving~(\ref{item:write:R_value_equals_contributions}), we will show that, at point $t$, there can be at most one process to which tag $g$ was transferred without the subsequent decrement, and that if such a process exists, then no object contributes to $R[g]$.
  Suppose then that at point $t' < t$, some process $p$ executes either \autoref{line:swap:swap},~\ref{line:sc2:first_cas}, or~\ref{line:cleanup:cas}, gets transferred the nonzero tag $g$ from $L_i$, and does not execute its next line until after point $t$.
  From \autoref{lem:write:locked_while_in_L}, $g$ remains locked throughout $[t', t]$.

  Suppose for the purpose of proving a contradiction that $L_j$ (possibly $L_j = L_i$) contributes to $R[g]$ at some point $t^\dagger \in (t', t)$; by \autoref{lem:write:contribute_while_in_L}, $L[j].\Tag = g$ at this point.
  (If a second process is transferred tag $g$ at some point in that interval, then $L[j].\Tag = g$ for some object $L_j$ immediately before the transfer, so that is also covered by this supposition.)
  Then there are points $t_1 < t_2 < t^\dagger$ such that at point $t_1$, some process $q$ locks tag $g$ in \autoref{line:swap:lock} or~\ref{line:sc2:lock}, and at point $t_2$, $q$ causes $L[j].\Tag$ to become $g$ in \autoref{line:swap:swap},~\ref{line:sc2:first_cas}, or~\ref{line:sc2:second_cas}.
  Since $g$ is locked for $L_i$ immediately before point $t'$, there is a point $t^\star < t' < t^\dagger$ at which $g$ becomes locked for $L_i$ by some process $r$.
  Since, from the code for \Swap{} and \SC{}, a process performs either a single \Swap{} on an object or at most one successful \CAS{} to set a \Tag field after locking a tag, it follows that $t^\star \neq t_1$.
  Whichever point is later between $t^\star$ and $t_1$ contradicts~(\ref{item:write:mutex_lock}), because $g$ becomes locked at the earlier of the two points, and remains locked until at least point $t$ which is later than both.

  In summary, each process other than one process with a transferred tag contributes at most one to $R[g]$, and there can be either one more contribution from an object, or one process with a transferred tag contributing two to $R[g]$.
  It follows that the total contribution to any element of $R$ is at most $n + 1$ in all cases, thus obtaining~(\ref{item:write:R_value_equals_contributions}) since the elements do not overflow.
\end{proof}

\begin{corollary}\label{cor:write:exclusive_tag}
  While tag $g$ is locked by process $p$ for $L_i$, only $p$ can cause $L[i].\Tag$ to change to $g$.
\end{corollary}
\begin{proof}
  Suppose that $L[i].\Tag$ changes to $g$ at point $t$ by process $q$, while $g$ is locked by process $p$.
  Since (by \autoref{def:write:lock}) only nonzero tags can be locked, the possible lines which can be executed at point $t$ are lines~\ref{line:swap:swap},~\ref{line:sc2:first_cas}, and~\ref{line:sc2:second_cas}.
  In each case, there is a point $t' < t$ at which $q$ breaks out of a loop, in \autoref{line:swap:lock} or~\ref{line:sc2:lock}.
  From \autoref{def:write:lock}, $q$ locks tag $g$ at point $t'$, and does not unlock $g$ until after point $t$.
  From \autoref{lem:write:R_value_and_mutex_lock}(\ref{item:write:mutex_lock}), $q = p$, proving the corollary.
\end{proof}

\begin{lemma}\label{lem:write:write_sc_changes_tag}
  If at point $t$ some process executes \autoref{line:swap:swap}, or a successful \CAS{} in \autoref{line:sc2:first_cas} or~\ref{line:sc2:second_cas}, then $L[i].\Tag$ changes from some tag $g$ to a new tag $g' \notin \set{g, 0}$ at point $t$.
\end{lemma}
This lemma is analogous to Lemma~18 of~\cite{BHW2026a-full}, but its proof uses the new concept of locking tags.
\begin{proof}
  Let $p$ be the process that executes \autoref{line:swap:swap}, or a successful \CAS{} in \autoref{line:sc2:first_cas} or~\ref{line:sc2:second_cas} at point $t$, and let $g'$ be the value of $L[i].\Tag$ immediately after that point.
  Since $p$ chooses $g'$ in \autoref{line:swap:choose_tag} or~\ref{line:sc2:choose_tag} from the set $\set{1, \dots, 4(k + 1)n}$, it follows that $g' \neq 0$, so it remains to show that $L[i].\Tag \neq g'$ immediately before $t$.

  Let $t' < t$ be the point at which $p$ breaks out of its loop in \autoref{line:swap:lock} or~\ref{line:sc2:lock}.
  Since $R[g] = 0$ immediately before $t'$, it follows from \autoref{lem:write:contribute_while_in_L} and~\autoref{lem:write:R_value_and_mutex_lock}(\ref{item:write:R_value_equals_contributions}) that $L[i] \neq g$ at $t'$.
  From \autoref{def:write:lock}, $g'$ is locked by $p$ at point $t'$, and remains locked until it decrements its contribution to $R[g]$, which by \autoref{def:write:contribute} occurs immediately after point $t$ (transferring its contribution to $L[i]$).
  Since $L[i].\Tag \neq g'$ at point $t'$, it follows from \autoref{cor:write:exclusive_tag} that $L[i].\Tag \neq g'$ throughout $[t', t)$, which concludes the proof.
\end{proof}

\subsection{Protecting Tags}

\begin{definition}[Protected and Strongly Protected Tags]\label{def:write:protect}
  A tag $g$ becomes \emph{protected} for LL/SC object $L_i$ at point $t$ by a process $p$ if either
  \begin{enumerate}[(a)]
    \item $p$ reads zero from $L[i].\Tag$ in \autoref{line:ll:first_read} or~\ref{line:ll:second_read} at point $t$, in which case $g = 0$, or\label{item:write:zero_protected}
    \item $p$ reads $g$ from $L[i].\Tag$ in \autoref{line:ll:second_read} or~\ref{line:cleanup:second_read} at point $t$, such that its local variable pair $(val, tag)$ (read from $L[i].\Tag$ in \autoref{line:ll:first_read} or~\ref{line:cleanup:first_read} respectively) equals the value of $L[i]$ at point $t$.\label{item:write:nonzero_protected}
  \end{enumerate}
  If a tag becomes protected for $L_i$ during an \LL{i} call, then it is also \emph{strongly protected} for $L_i$ by $p$.
  In that case, the protection and strong protection by process $p$ end when the calling process next invokes \CL{i}.
  Otherwise, the tag becomes protected for $L_i$ during a \cleanup{i} call, in which case it is not strongly protected by process $p$, and the protection by $p$ ends when that call responds.

  If a tag is protected or strongly protected by any process at point $t$ for $L_i$, we also simply refer to it as protected or strongly protected for $L_i$, without referring to the protecting process.
\end{definition}
The only difference compared to Definition 9~of~\cite{BHW2026a-full} is that protecting tags in the \CL{} method moves to the \cleanup{} helper method.

\begin{lemma}\label{lem:write:protected_activity}
  If some tag is strongly protected for $L_i$ at point $t$, then $C[i] \neq 0$ at point $t$.
\end{lemma}
This lemma corresponds to Lemma~10 of~\cite{BHW2026a-full}, and its proof is nearly identical.
\begin{proof}
  From \autoref{def:write:protect}, a tag can only become strongly protected for $L_i$ by a process $p$ in an \LL{i} call in \autoref{line:ll:first_read} or~\ref{line:ll:second_read}, and both lines happen after $p$ increments $C[i]$ in \autoref{line:ll:inc_activity} and before its next invocation of \CL{i}.
  Since only \CL{i} can decrement $C[i]$ (in \autoref{line:cl2:dec_activity}), it follows that $p$'s strong protection of $g$ for $L_i$ ends before it decrements $C[i]$.
  From \autoref{def:write:contribute}, $p$ contributes to $C[i]$ throughout its strong protection of $g$.
  The lemma then follows from \autoref{lem:write:process_contribution}(\ref{item:write:contribution:C}) and \autoref{lem:write:C_value_equals_contributions}.
\end{proof}

\begin{lemma}\label{lem:write:idle_protection}
  If a process $p$ has no pending method call on $L_i$ at point $t$, then $p$ strongly protects tag $g$ for $L_i$ if and only if $d[i].\Tag = g$ for $p$'s dictionary $d$ at point $t$.
\end{lemma}
This corresponds to Lemma~11 of~\cite{BHW2026a-full}, and the proof is essentially identical.
\begin{proof}
  We first note that the only case in which $p$ can protect a tag for $L_i$ while not having a pending method call on $L_i$, by \autoref{def:write:protect}, is if it protected the tag as part of an \LL{i} call.
  In particular, this implies that $p$ also strongly protects the tag.
  Moreover, only \LL{i} can set $d[i]$ to a non-$\bot$ value (in lines~\ref{line:ll:set_di} and~\ref{line:ll:reset_tag}) and only \CL{i} can reset $d[i]$ to $\bot$ (in \autoref{line:cl2:clear_di}).
  An invocation of \CL{i} ends $p$'s (strong) protection of any tag for $L_i$ by \autoref{def:write:protect}, and, from \autoref{lem:write:cl_clears_di}, $d[i] = \bot$ when \CL{i} responds.
  It therefore remains to show that an \LL{i} call strongly protects tag $g$ for $L_i$ by its response if and only if $d[i].\Tag = g$ at that call's response.

  To show this, we consider the five possible response points from an \LL{i} call by process $p$.
  Suppose by induction that the lemma is true before the call begins (the base case holds since $d[i] = \bot$ and processes protect no tags at $t = 0$).
  If $p$ protects some tag $g$ for $L_i$ just before invocation of the \LL{i} call, then $d[i].\Tag = g$ before invocation, and therefore $p$ invokes \CL{i} in \autoref{line:ll:reset_link}.
  Thus, after \autoref{line:ll:reset_link}, $p$ does not protect any tag for $L_i$, and $d[i] = \bot$.

  If $p$ responds in \autoref{line:ll:first_zero_tag} or~\ref{line:ll:second_zero_tag}, then $p$ reads zero from $L[i].\Tag$ in \autoref{line:ll:first_read} or~\ref{line:ll:second_read} respectively.
  By \autoref{def:write:protect}(\ref{item:write:zero_protected}), $p$ strongly protects tag zero for $L_i$ and sets $d[i].\Tag$ to zero in \autoref{line:ll:set_di} or~\ref{line:ll:reset_tag} respectively before it responds.

  If $p$ responds in \autoref{line:ll:matching_tag}, then $p$ re-reads its local variables $(val, tag)$ from $L[i]$ in \autoref{line:ll:second_read}after having set $d[i]$ to $(val, tag)$ in \autoref{line:ll:set_di} and not changing it afterwards.
  By \autoref{def:write:protect}(\ref{item:write:nonzero_protected}), $p$ strongly protects $d[i].\Tag$ in \autoref{line:ll:second_read}, at which point $L[i] = d[i]$.

  Otherwise, $p$ responds in \autoref{line:ll:final_return}, it which case the \CL{i} call invoked in \autoref{line:ll:clear} resets $d[i]$ to $\bot$ and ends any protection of a tag by $p$ for $L_i$.
\end{proof}

\begin{lemma}\label{lem:write:protect_during_sc}
  Suppose that a successful \CAS{} takes place on $L[i]$ at point $t$, such that $L[i].\Tag \neq 0$ immediately after $t$.
  At point $t$, some process $p$ executes \autoref{line:sc2:first_cas} or~\ref{line:sc2:second_cas} of an \SC{i, \cdot} call.
  Process $p$ strongly protects some tag $g^\star$ for $L_i$ from the invocation of its call until at least point $t$. %
\end{lemma}
This is a slightly simpler version of Lemma~12 from~\cite{BHW2026a-full}.
Note that the lemma phrasing supposes that a \CAS{} operation specifically takes place at point $t$, which rules out that a \Swap{} in \autoref{line:swap:swap} takes place instead.
\begin{proof}
  Since only lines~\ref{line:sc2:first_cas} and~\ref{line:sc2:second_cas} are \CAS{} operations which can change $L[i].\Tag$ to a nonzero value, it follows that point $t$ is the point that some process $p$ executes one of the two lines.
  Since $p$ does not return in \autoref{line:sc2:early_return}, by \autoref{lem:write:idle_protection}, $p$ strongly protects $g^\star = d[i].\Tag$ for $L_i$ when it invokes \SC{i, \cdot}.
  Moreover, since $p$ does not invoke \CL{i} in between invoking \SC{i, \cdot} and its successful \CAS{} at point $t$, by \autoref{def:write:protect} $g^\star$ remains strongly protected for $L_i$ until at least point $t$.
\end{proof}

\subsection{History Independence}
Having established the useful properties of our implementation, we can now prove our results, starting with the QHI-preserving property.
This result, proven in \autoref{thm:write:history_independence}, is the same statement as Theorem~15 of~\cite{BHW2026a-full}, but with more work needed to prove it.
\begin{lemma}\label{lem:write:unprotected}
  If no process has a pending method call on $L_i$ and no process strongly protects a tag for $L_i$ at point $t$, then $C[i] = 0$ at point $t$.
  If no process has a pending method call on any object and no process strongly protects any tag at point $t$, then $R[g] = 0$ for all $g \in \set{1, \dots, 4(k + 1)n}$.
\end{lemma}
\begin{proof}
  Suppose that no process has a pending method call on $L_i$ and no process strongly protects any tag for $L_i$ at point $t$.
  If $d[i] \neq \bot$ for some process, then $d[i].\Tag = g$ for some tag $g \in \set{0, \dots, 4(k + 1)n}$.
  From \autoref{lem:write:idle_protection}, $p$ would strongly protect $g$ at point $t$: thus, $d[i] = \bot$ for all processes at point $t$.
  \autoref{lem:write:process_contribution}(\ref{item:write:idle_contribution:C}) and \autoref{lem:write:C_value_equals_contributions} then imply that $C[i] = 0$ at point $t$.
  If no process has any pending method call or strongly protects any tag at point $t$, then \autoref{lem:write:process_contribution}(\ref{item:write:idle_contribution:R}) and \autoref{lem:write:R_value_and_mutex_lock}(\ref{item:write:R_value_equals_contributions}) imply that $R[g] = 0$ for all $g$.
\end{proof}

\begin{lemma}\label{lem:write:quiescent_zero_tag}
  If at point $t$ no process has a pending method call on $L_i$, except possibly a pending next step of \autoref{line:cleanup:dec_prev} or~\ref{line:cleanup:final_return}, and no process strongly protects a tag for $L_i$, then $L[i].\Tag = 0$ at point $t$.
\end{lemma}
This lemma corresponds to Lemma~14 of~\cite{BHW2026a-full}, but with more cases to consider.
\begin{proof}
  The lemma is true for $t = 0$, since $L[i].\Tag$ is initialized to zero for each $i \in \set{1, \dots, m}$.
  The \Read{i} method cannot be pending and does not affect which tags the calling process protects or the value of $L[i].\Tag$, so we need not consider it for this lemma.
  Similarly, by \autoref{lem:write:idle_protection}, we have that method calls which respond in either \autoref{line:sc2:early_return},~\ref{line:cl2:early_return}, or~\ref{line:vl:early_return} do not affect which tags the calling process strongly protects or the value of $L[i].\Tag$.
  Method calls on objects $L_j \neq L_i$ also do not affect whether the calling process strongly protects a tag for $L_i$ of the value of $L[i].\Tag$.
  Note that all the method calls mentioned so far are either not for $L_i$, or consist of at most one shared memory step before returning.
  Therefore, none of them change whether a process has a pending method call on $L_i$.
  We can therefore assume without loss of generality that point $t$ is either the response point of a method not covered by this paragraph, or (immediately after) an execution of \autoref{line:cleanup:cas}.

  By inspection of the code and by \autoref{def:write:protect} and \autoref{lem:write:idle_protection}, the remaining method calls for which the calling process responds without strongly protecting a tag for $L_i$ at the response point are exactly
  \begin{enumerate}[(a)]
    \item an \LL{i} call that responds in \autoref{line:ll:final_return},
    \item an \SC{i, \cdot} call that responds in \autoref{line:sc2:return},
    \item a \CL{i} call that responds after \autoref{line:cl2:early_return},
    \item a \VL{i} call that responds in \autoref{line:vl:return_false}, and
    \item a \Swap{i, \cdot} call.
  \end{enumerate}
  The cases of responding in either \autoref{line:ll:final_return},~\ref{line:sc2:return}, or~\ref{line:vl:return_false} all immediately follow a \CL{i} call.
  As such, we can further assume that point $t$ is either the response point or an execution of \autoref{line:cleanup:cas} from a \CL{i} or a \Swap{i, \cdot} call $c$.

  By \autoref{lem:write:unprotected}, $C[i] = 0$ at point $t$ if it is a response point.
  Since lines~\ref{line:cleanup:dec_prev} and~\ref{line:cleanup:final_return} do not affect $C[i]$, and in all cases there are no shared memory operations after \autoref{line:cleanup:final_return} until the calling process becomes idle, it also follows that $C[i] = 0$ if $t$ is an execution of \autoref{line:cleanup:cas}.
  Since call $c$ does not respond in \autoref{line:cl2:early_return}, either it decrements $C[i]$ in \autoref{line:cl2:dec_activity}, or it is a \Swap{} call.
  Therefore, there is a point before $t$ at which either $C[i]$ is decremented to zero, or \autoref{line:swap:swap} of a \Swap{i, \cdot} call takes place.
  (Note from the code that neither can take place at point $t$ itself.)
  Let $t'$ be the last such point before point $t$.
  Since only \autoref{line:cl2:dec_activity} decrements $C[i]$, it follows that either some \CL{i} call $c'$ (possibly $c' = c$) decrements $C[i]$ to zero at point $t'$, or a \Swap{i, \cdot} call $c'$ executes \autoref{line:swap:swap} at point $t'$.

  Let $p$ be the process executing call $c'$.
  By definition of $t'$, we also have that no \Swap{i, \cdot} call executes \autoref{line:swap:swap} in the interval $(t', t]$, and $C[i] = 0$ throughout the interval.
  If $c'$ is a \CL{i} call, then $p$ sees a $1$ from the \FAD{} in \autoref{line:cl2:dec_activity}, and then invokes \cleanup{i}.
  Otherwise, $c'$ is a \Swap{i, \cdot} call, and so it reads $C[i] = 0$ when it executes \autoref{line:swap:confirm_quiescence} in $(t', t]$, and therefore also invokes \cleanup{i}.
  Since no process has a pending method call on $L_i$ at point $t$ other than possibly lines~\ref{line:cleanup:dec_prev} or~\ref{line:cleanup:final_return}, $p$'s call has executed \autoref{line:cleanup:cas} or responded by point $t$.

  Since $C[i] = 0$ throughout the interval $(t', t]$, by \autoref{lem:write:protected_activity}, we have that no tag is strongly protected for $L_i$ throughout $(t', t]$.
  Therefore, from \autoref{lem:write:protect_during_sc}, no successful \CAS{} takes place on $L[i]$ in $(t', t]$ that sets $L[i].\Tag$ to a nonzero value.
  Since the only other line which can change $L[i].\Tag$ to a nonzero value is \autoref{line:swap:swap}, and this line is not executed in $(t', t]$, it follows that $L[i].\Tag$ does not change to a nonzero value in that interval.
  In particular, if $L[i].\Tag$ changes at all in the interval, it only does so from a successful \CAS{} in \autoref{line:cleanup:cas}, which changes $L[i].\Tag$ to zero without changing $L[i].\val$.
  It thus suffices to show that $L[i].\Tag = 0$ at some point during $p$'s \cleanup{i} call.

  If $p$ responds from its \cleanup{i} call in \autoref{line:cleanup:zero_tag_return}, or if the if-condition in \autoref{line:cleanup:second_read} is \False, then $L[i].\Tag$ changes at some point in $(t', t]$, and therefore $L[i].\Tag$ changes to zero at that point, proving the lemma.
  Otherwise, we have that $p$ reads zero from $C[i]$ in \autoref{line:cleanup:confirm_quiescence}, and therefore executes the \CAS{} in \autoref{line:cleanup:cas}.
  If the \CAS{} succeeds, then $L[i].\Tag = 0$ immediately after it, and therefore by point $t$, proving the lemma.
  Otherwise, $L[i]$ changes at some point after $t'$ and before the \CAS{} attempt, and so we again have that $L[i].\Tag = 0$ at that point.
\end{proof}

Our first main theorem establishes that in a settled state (defined in \autoref{sec:intro}), all shared memory apart from LL/SC values stored in $\val$ fields are set to their default values of zero, implying that our implementation preserves quiescent history independence.
\begin{theorem}\label{thm:write:history_independence}
  If every process is idle and has no outstanding \LL{} operation on any LL/SC object at point $t$, then all values of $C$, $R$, and all \Tag fields of $L$ are zero at point $t$.
\end{theorem}
This theorem corresponds to Theorem~15 of~\cite{BHW2026a-full}.
\begin{proof}
  From \autoref{lem:write:nonbot_pending_link}, all values of $d$ are $\bot$ for every process at point $t$.
  From \autoref{lem:write:idle_protection}, no process strongly protects any tag at point $t$.
  The theorem then follows from Lemmas~\ref{lem:write:unprotected} and~\ref{lem:write:quiescent_zero_tag}.
\end{proof}

\subsection{Expected Step Complexity}
The core idea to prove expected step complexity is the same as in~\cite{BHW2026a-full}, but the main additional difficulty compared to that paper is in keeping track of contributions in \autoref{sec:analysis:contributions} and \autoref{lem:write:max_total_contribution}.

\begin{lemma}\label{lem:write:max_total_contribution}
  At all points in the execution, the total contribution to all elements of $R$ is at most $2(k + 1)n$.
\end{lemma}
\begin{proof}
  From \autoref{lem:write:process_contribution}(\ref{item:write:contribution:R}), each process contributes at most $k + 2$ in total to the elements of $R$ at any given time.

  From Lemmas~\ref{lem:write:contribute_while_in_L} and~\ref{lem:write:quiescent_zero_tag}, for each LL/SC object $L_i$ which contributes one to some element of $R$, there exists a process that either
  \begin{itemize}
    \item has a pending method call on $L_i$, other than a pending next step of \autoref{line:cleanup:dec_prev} or~\ref{line:cleanup:final_return}, or
    \item strongly protects some tag for $L_i$.
  \end{itemize}
  In the case of a process without a pending method call on $L_i$ which strongly protects some tag for $L_i$, by \autoref{lem:write:idle_protection} $d[i] \neq \bot$ for that process, and by \autoref{lem:write:nonbot_pending_link} the process has an outstanding \LL{i}.
  As required in~\ref{sec:write:pending_links}, each process has at most $k$ outstanding \LL{} operations on objects for which it does not have a pending method call.
  Thus, each process can be associated with at most $k + 1$ contributions to $R$ from LL/SC objects.

  However, from \autoref{lem:write:nonbot_pending_link}, \autoref{def:write:contribute}, \autoref{lem:write:process_contribution}(\ref{item:write:idle_contribution:R}), and inspection of the code, the only situation in which a process can contribute exactly $k + 2$ to the elements of $R$ is if
  \begin{itemize}
    \item the process has $k$ outstanding \LL{i} operations, and
    \item its last step was a successful \CAS{} in \autoref{line:cleanup:cas}, making its pending next step of \autoref{line:cleanup:dec_prev}.
  \end{itemize}
  This scenario is mutually exclusive with that of an LL/SC object $i$ contributing to an element of $R$ due to that same process having a pending method call on $L_i$, since it cannot have \autoref{line:cleanup:dec_prev} as its pending next step.
  As such, each process can be associated with at most $k + 2 + k + 1 - 1 = 2k + 2$ total contributions to $R$.
  Thus, the total contribution to $R$ is at most $2(k + 1)n$.
\end{proof}

\begin{theorem}[Expected Step Complexity]\label{thm:write:complexity}
  Each method other than \Swap{} and \SC{} has a constant bound on its number of steps.
  The expected step complexity of the \Swap{} and \SC{} methods is constant under a weak adaptive adversary.
\end{theorem}
This theorem corresponds to Theorem 7 of~\cite{BHW2026a-full}.
\begin{proof}
  The first part is immediate from the fact that the only loops in the implementation are lines~\ref{line:swap:repeat} to~\ref{line:swap:lock} of the \Swap{} method and lines~\ref{line:sc2:repeat} to~\ref{line:sc2:lock} of the \SC{} method.

  For the second part, recall that under the weak adaptive adversary, a random choice by process $p$ is immediately followed by its next shared memory step.
  Thus, when $p$ chooses a random tag $g$ in \autoref{line:swap:choose_tag} or~\ref{line:sc2:choose_tag}, it then immediately attempts a \CAS{} on $R[g]$ in \autoref{line:swap:lock} or~\ref{line:sc2:lock}, respectively, without any other process taking a step in between the two respective lines.
  In particular, the value of $R[g]$ does not change in between the two lines.
  Therefore, $p$'s \CAS{} on $R[g]$ from zero to one in \autoref{line:swap:lock} or~\ref{line:sc2:lock} succeeds if and only if $R[g] = 0$ when $p$ executes \autoref{line:swap:choose_tag} or~\ref{line:sc2:choose_tag}, respectively.
  Thus, $p$ breaks out of its loop as soon as it chooses a tag $g$ such that $R[g] = 0$ when $p$ executes \autoref{line:swap:choose_tag} or~\ref{line:sc2:choose_tag}, in which case it sets $R[g]$ to one upon exiting the loop.
  Since this random choice is uniform over $\set{1, \dots, 4(k + 1)n}$ and independent, it suffices to show that, at any given time, at least a constant fraction of the indices in $R$ have a value of zero.

  From \autoref{lem:write:max_total_contribution}, the total contribution to all elements of $R$ is at most $2(k + 1)n$, and from \autoref{cor:write:non-negative_contribution}, all contributions are non-negative.
  Thus, at any given time, at least $2(k + 1)n$ of $R$'s $4(k + 1)n$ elements have zero contribution and hence, by \autoref{lem:write:R_value_and_mutex_lock}(\ref{item:write:R_value_equals_contributions}), a value of zero, which completes the proof.
\end{proof}

\subsection{Linearizability}\label{sec:linearizability}
We conclude the analysis by proving in \autoref{thm:write:linearizable} that the algorithm is a linearizable implementation of the sequential specification from \autoref{sec:write:seq_spec}.
There are more cases to consider compared to~\cite{BHW2026a-full} due to the interaction between the \Swap{} and \SC{} methods, but the new concept of locked tags makes some arguments a little simpler.

\begin{lemma}\label{lem:write:protected_refcount}
  If tag $g \neq 0$ is protected for some object at point $t$, then $R[g] \neq 0$ at point $t$.
\end{lemma}
This corresponds to Lemma~16 of~\cite{BHW2026a-full}, with a slightly modified proof.
\begin{proof}
  Suppose that process $p$ protects tag $g \neq 0$ for some LL/SC object $L_i$ at point $t^\star$.
  Observe from \autoref{def:write:protect}(\ref{item:write:nonzero_protected}) that when a nonzero tag $g$ is protected in \autoref{line:ll:second_read} or~\ref{line:cleanup:second_read}, the process $p$ protecting it has incremented $R[g]$ (in line \autoref{line:ll:inc_refcount} or~\ref{line:cleanup:inc_refcount} respectively) and not yet decremented $R[g]$.
  From \autoref{def:write:contribute}, $p$ therefore contributes a nonzero value to $R[g]$ at point $t^\star$.

  In the case of \autoref{line:ll:second_read}, by \autoref{def:write:protect}(\ref{item:write:nonzero_protected}), $L[i]$ matches $p$'s variable pair $(val, tag)$ at point $t^\star$ (so $L[i].\Tag = tag = g$); therefore, $p$ returns in \autoref{line:ll:matching_tag}.
  If $p$ decrements its contribution by transferring it to an LL/SC object in either \autoref{line:swap:swap},~\ref{line:sc2:first_cas}, or~\ref{line:sc2:second_cas}, then that decrement cancels out an increment in either \autoref{line:swap:lock} or~\ref{line:sc2:lock} of the same method call.
  Also, if $p$ decrements its contribution in either \autoref{line:swap:dec_prev},~\ref{line:sc2:dec_prev}, or~\ref{line:cleanup:dec_prev}, then a contribution from an LL/SC object would have been previously transferred to $p$ in either \autoref{line:swap:swap},~\ref{line:sc2:first_cas}, or~\ref{line:cleanup:cas}.
  Thus, $p$'s contribution to $R[g]$ does not decrement to zero until at least the next point after $t^\star$ at which it decrements the value of $R[g]$ within a \CL{i} call, in which case $g$ ceases to be protected for $L_i$ at that call's invocation, or in \autoref{line:ll:dec_refcount} of a subsequent invocation of \LL{i}.
  However, since only \CL{i} can reset $d[i]$ to $\bot$ (in \autoref{line:cl2:clear_di}), if $p$ invokes \LL{i} after point $t^\star$ without invoking \CL{i} in between, then it invokes \CL{i} in \autoref{line:ll:reset_link}.
  This makes $g$ no longer protected for $L_i$ before $p$ decrements $R[g]$ in \autoref{line:ll:dec_refcount}.

  In the case of \autoref{line:cleanup:second_read}, $p$ decrements $R[g]$ in \autoref{line:cleanup:final_return}, and possibly also in \autoref{line:cleanup:dec_prev}.
  As mentioned earlier, if $p$ decrements $R[g]$ in \autoref{line:cleanup:dec_prev}, then it also increments its contribution to $R[g]$ via a transfer from $L_i$ in \autoref{line:cleanup:cas}.
  Hence, $p$'s contribution can only decrement to zero in \autoref{line:cleanup:final_return}, at which point $p$'s protection of $g$ ends.

  Therefore, $p$ does not decrement its contribution to $R[g]$ after point $t^\star$ until after $g$ is no longer protected for $L_i$.
  The lemma then follows from \autoref{lem:write:process_contribution}(\ref{item:write:contribution:R}) and \autoref{lem:write:R_value_and_mutex_lock}(\ref{item:write:R_value_equals_contributions}).
\end{proof}

\begin{lemma}[Protection Invariant]\label{lem:write:invariant}
  Suppose that tag $g$ is protected for $L_i$ at point $t$. Then
  \begin{enumerate}[(a)]
    \item $L[i].\Tag$ does not change from a value different from $g$ to $g$ at point $t$, and\label{item:write:invariant:protection}
    \item if $g$ is \emph{strongly} protected for $L_i$ at point $t$, then $L[i].\Tag$ also does not change from a value different from $g$ to zero at point $t$.\label{item:write:invariant:strong_protection}
  \end{enumerate}
\end{lemma}
This lemma corresponds to Lemma~17 of~\cite{BHW2026a-full}, but the proof of~(\ref{item:write:invariant:protection}) uses a different technique, using the new concept of locked tags.
\begin{proof}
  We prove this lemma by contradiction.
  Suppose that the invariant is not true, and let $t$ be the first point in an execution at which it fails.
  Thus, at point $t$, there exists a tag $g$ that is protected for $L_i$, and $L[i].\Tag$ changes from a tag $g' \neq g$ to $g$, or to zero if $g$ is strongly protected for $L_i$.

  Suppose first that~(\ref{item:write:invariant:strong_protection}) fails at point $t$, and therefore that $g$ is strongly protected for $L_i$ at point $t$, but that $L[i].\Tag$ changes from some tag $g' \neq g$ to zero at that point.
  Only \autoref{line:cleanup:cas} can change $L[i].\Tag$ to zero, since the only other lines which can change $L[i].\Tag$, namely lines~\ref{line:swap:swap},~\ref{line:sc2:first_cas}, and~\ref{line:sc2:second_cas}, choose a tag that is nonzero (from lines~\ref{line:swap:choose_tag} and~\ref{line:sc2:choose_tag}).
  Thus, there is a process $p$ which executes a successful \CAS{} in \autoref{line:cleanup:cas} at point $t$, changing $L[i].\Tag$ from $g'$ ($p$'s local value of $tag$) to zero.
  The if conditions that $p$ checks in lines~\ref{line:cleanup:second_read} and~\ref{line:cleanup:confirm_quiescence} immediately before its \CAS{} at point $t$ both return true, since $p$ executes \autoref{line:cleanup:cas}.
  Therefore, there exist points $t_L < t_C < t$, such that $L[i].\Tag = g'$ at point $t_L$, and $C[i] = 0$ at point $t_C$.

  From \autoref{def:write:protect}(\ref{item:write:nonzero_protected}), $g'$ becomes protected for $L_i$ at point $t_L$, and it remains protected until $p$ executes \autoref{line:cleanup:final_return}, which is after point $t$.
  From \autoref{lem:write:protected_activity}, $g$ is not strongly protected for $L_i$ at point $t_C$.
  Since by assumption $g$ is strongly protected for $L_i$ at point $t > t_C$, $g$ becomes strongly protected for $L_i$ at some point $t^\star \in (t_C, t)$.
  By \autoref{def:write:protect}, $L[i].\Tag = g$ at point $t^\star$.
  Since $L[i].\Tag = g' \neq g$ just before point $t > t^\star$, there is a point $t^\dagger \in (t^\star, t)$ at which $L[i].\Tag$ changes from some value different from $g'$ (possibly $g$) to $g'$.
  But since $g'$ is protected for $L_i$ throughout $(t_L,t]$ (recall that $t_L < t_C < t^\star < t^\dagger < t$), point $t^\dagger$ violates~(\ref{item:write:invariant:protection}), contradicting that $t$ is the first point at which the invariant fails.

  It remains to derive a contradiction for the case where~(\ref{item:write:invariant:protection}) fails at point $t$, and therefore that $L[i].\Tag$ changes from some tag $g' \neq g$ to $g$ at point $t$, while $g$ is protected for $L_i$.
  Note that if $g = 0$, then it cannot become protected in \autoref{line:cleanup:second_read}, as the \cleanup{} method would exit in \autoref{line:cleanup:zero_tag_return}.
  Thus, if $g = 0$ is protected then it is protected during an \LL{} call, and therefore it is also strongly protected, so the contradiction follows from that just derived for~(\ref{item:write:invariant:strong_protection}).
  We can therefore assume without loss of generality that $g \neq 0$.

  Since the only lines that can change $L[i].\Tag$ to a nonzero value are lines~\ref{line:swap:swap},~\ref{line:sc2:first_cas}, and~\ref{line:sc2:second_cas}, there is a point $t' < t$ at which $p$ increments $R[g]$ from zero to one, in \autoref{line:swap:lock} or~\ref{line:sc2:lock}.
  By \autoref{def:write:lock}, $g$ is locked by $p$ throughout the interval $(t', t]$.
  Since $R[g] = 0$ immediately before point $t'$, by \autoref{lem:write:protected_refcount} we have that $g$ is not yet protected immediately before $t'$.
  Since, by \autoref{def:write:protect}, lines~\ref{line:swap:lock} and~\ref{line:sc2:lock} do not cause a tag to become protected, $g$ is also not protected at point $t'$, and so $g$ becomes protected at some point $t^\star \in (t', t)$.
  By \autoref{def:write:protect}, $L[i].\Tag = g$ at point $t^\star$.
  Since $R[g] = 0$ immediately before $t'$, by \autoref{lem:write:contribute_while_in_L} and \autoref{lem:write:R_value_and_mutex_lock}(\ref{item:write:R_value_equals_contributions}) we also have that $L[i].\Tag \neq g$ at point $t'$ (recall that $L[i]$ does not change at point $t'$).
  Therefore, there is a point $t^\dagger \in (t', t^\star)$ at which $L[i].\Tag$ changes from some value different from $g$ to $g$.
  The process taking the step at point $t^\dagger$ is not $p$, since $p$ does not change the value of $L[i]$ in between points $t'$ and $t$.
  Since $g$ is locked by $p$ at point $t^\dagger \in (t', t]$, this violates \autoref{cor:write:exclusive_tag}, which is a contradiction.
\end{proof}

\begin{definition}[Function $lin$]\label{def:write:lin}
  We define a function $lin(c)$ from method calls $c$ in an execution to points in time, and later (in \autoref{thm:write:linearizable}) prove that they are correct linearization points for the implementation.

  If $c$ is an \LL{} call, then $lin(c)$ depends on which line the method responds from.
  If it responds in \autoref{line:ll:first_zero_tag} or~\ref{line:ll:final_return}, then $lin(c)$ occurs at \autoref{line:ll:first_read}.
  If it responds in \autoref{line:ll:matching_tag} or~\ref{line:ll:second_zero_tag}, then $lin(c)$ occurs at \autoref{line:ll:second_read}.

  If $c$ is a \Swap{} call, then $lin(c)$ occurs at \autoref{line:swap:swap}.

  If $c$ is an \SC{} call, then $lin(c)$ occurs at its invocation if it returns in \autoref{line:sc2:early_return}, and otherwise $lin(c)$ is the point of the last \CAS{} operation on $L$.
  That is, if the \CAS{} in \autoref{line:sc2:first_cas} succeeds, then $lin(c)$ occurs at \autoref{line:sc2:first_cas}, and otherwise $lin(c)$ occurs at \autoref{line:sc2:second_cas}.

  If $c$ is a \CL{} call, then $lin(c)$ occurs at its invocation.

  If $c$ is a \Read{} call, then $lin(c)$ occurs at \autoref{line:read} (its only step).

  If $c$ is a \VL{} call, then $lin(c)$ occurs at its invocation if it returns in \autoref{line:vl:early_return}, and otherwise $lin(c)$ occurs at \autoref{line:vl:read}.

  For an execution with pending method calls, $lin(c)$ is undefined for pending \LL{} calls, and for other method calls for which the line corresponding to $lin(c)$ has not yet been executed.
\end{definition}
The only new addition relative to Definition~19 from~\cite{BHW2026a-full} is the \Swap{} method, for which $lin(c)$ maps to the intuitive point of \autoref{line:swap:swap}.

\begin{lemma}\label{lem:write:successful}
  Suppose that $c$ is either an \SC{i, u} or \VL{i} call by process $p$.
  The call returns \True, and changes $L[i].\val$ to $u$ at point $lin(c)$, in the case of \SC{i, u}, if and only if
  \begin{enumerate}[(a)]
    \item some tag becomes strongly protected for $L_i$ by $p$ at some point $t < lin(c)$, and remains strongly protected by $p$ throughout $(t, lin(c)]$,\label{item:write:successful:strong_protection}
    \item no \Swap{i, \cdot} call executes the \Swap{} hardware instruction in \autoref{line:swap:swap} in the interval $(t, lin(c))$.\label{item:write:successful:no_write}
    \item no \SC{i, \cdot} call has a successful \CAS{} in \autoref{line:sc2:first_cas} or~\ref{line:sc2:second_cas} in the interval $(t, lin(c))$, and\label{item:write:successful:no_others}
  \end{enumerate}
\end{lemma}
This lemma corresponds to Lemma~20 of~\cite{BHW2026a-full}, but here we have the new condition~(\ref{item:write:successful:no_write}).
The proof is mostly unchanged from that of~\cite{BHW2026a-full}.
\begin{proof}
  We first prove the forward direction, so we first assume that $c$ returns \True.
  Since $c$ returns \True, it responds in either \autoref{line:vl:return_true} or~\ref{line:sc2:return}.
  Since in particular it does not return in \autoref{line:vl:early_return} or~\ref{line:sc2:early_return}, it means that $d[i] \neq \bot$ when $p$ invokes $c$.
  By \autoref{lem:write:idle_protection}, $d[i].\Tag$ is strongly protected at $c$'s invocation, starting from some previous point $t < lin(c)$, and by \autoref{def:write:protect} it remains protected until after $lin(c)$.
  Thus,~(\ref{item:write:successful:strong_protection}) is satisfied for some point $t < lin(c)$.

  Let $g^\star$ be the tag that $p$ strongly protects at point $t$, i.e., its value of $d[i].\Tag$ in call $c$.
  Suppose for the purpose of proving a contradiction that either~(\ref{item:write:successful:no_write}) or~(\ref{item:write:successful:no_others}) is false, so either \autoref{line:swap:swap} or a successful \CAS{} in either \autoref{line:sc2:first_cas} or~\ref{line:sc2:second_cas} takes place in $(t, lin(c))$, and let $t'$ be the first such point.
  Note that the only other operation which can change the value of $L[i]$ happens in \autoref{line:cleanup:cas}, which also sets $L[i].\Tag$ to zero.
  Since, by \autoref{def:write:protect}, $L[i].\Tag = g^\star$ at point $t$, it follows that $L[i].\Tag$ is either $g^\star$ or zero immediately before point $t'$.
  By \autoref{lem:write:write_sc_changes_tag}, $L[i].\Tag$ changes to a different and nonzero value at point $t'$, and by \autoref{lem:write:invariant}, $L[i].\Tag$ does not change to $g^\star$ at point $t'$ since $g^\star$ is protected at that point.
  Thus, $L[i].\Tag$ changes to some nonzero tag distinct from $g^\star$ at point $t'$.
  Applying \autoref{lem:write:invariant} to each change to $L[i].\Tag$ in $(t', lin(c))$ (if any), we obtain that $L[i].\Tag$ can only change to other nonzero values distinct from $g^\star$ throughout the interval.
  But, since we assume that $c$ returns \True, and $lin(c)$ occurs at either \autoref{line:sc2:first_cas},~\ref{line:sc2:second_cas}, or~\ref{line:vl:read}, it follows from the code that $L[i].\Tag \in \set{g^\star, 0}$ immediately before $lin(c)$, which is a contradiction.
  Thus,~(\ref{item:write:successful:no_others}) and~(\ref{item:write:successful:no_write}) follow, finishing the proof for the forward direction.

  It remains to show the reverse direction, so we assume that~(\ref{item:write:successful:strong_protection}),~(\ref{item:write:successful:no_others}), and~(\ref{item:write:successful:no_write}) each hold for $c$.
  Like the forward case, let $g^\star$ be the tag that $p$ strongly protects throughout $(t, lin(c)]$, and we obtain from \autoref{def:write:protect} that $L[i].\Tag = g^\star$ at point $t$.
  From \autoref{lem:write:idle_protection},
  $p$'s value of $d[i].\Tag$ is $g^\star$ at $c$'s invocation, and in particular $p$ does not return in \autoref{line:vl:early_return} or~\ref{line:sc2:early_return}.
  Since no step which can change $L[i].\Tag$ to a nonzero value (i.e., \autoref{line:swap:swap} or a successful \CAS{} in either \autoref{line:sc2:first_cas} or~\ref{line:sc2:second_cas}) takes place throughout $(t, lin(c))$, it follows that $L[i].\Tag \in \set{g^\star, 0}$ immediately before point $lin(c)$.
  If $c$ is a \VL{i} call, then the call returns \True because $g^\star$ is $p$'s value of $d[i].\Tag$.
  If $c$ is an \SC{i, \cdot} call and its \CAS{} in \autoref{line:sc2:first_cas} succeeds, then the call returns \True.
  Otherwise, the \CAS{} in \autoref{line:sc2:first_cas} fails, so $p$ performs a second \CAS{} in \autoref{line:sc2:second_cas}, and $lin(c)$ is the point of that second \CAS{}.
  Since the only possible transition of $L[i].\Tag$'s value in $(t, lin(c))$ is from $g^\star$ to zero, it follows that $L[i].\Tag = 0$ when $p$ executes \autoref{line:sc2:first_cas}, and therefore remains at zero immediately before $lin(c)$, making the second \CAS{} successful.
\end{proof}

\begin{lemma}\label{lem:write:ll_strong_protection}
  Suppose that $c$ is an \LL{i} call by process $p$, such that $lin(c) = t$ and that the call responds at point $t'$.
  If some tag becomes strongly protected for $L_i$ by $p$ during the call, then it becomes strongly protected at point $t$.
  Otherwise, some process $q \neq p$ executes a call $c'$ which is either a \Swap{i, \cdot} call or an \SC{i, \cdot} call with a successful \CAS{} at point $lin(c')$, such that $t < lin(c') < t'$.
\end{lemma}
This corresponds to Lemma~21 of~\cite{BHW2026a-full}, with the only difference that an overlapping \Swap{} operation can also cause an \LL{} call to not strongly protect a tag.
\begin{proof}
  Combining Definitions~\ref{def:write:protect} and~\ref{def:write:lin}, an \LL{} call $c$ that responds in \autoref{line:ll:first_zero_tag},~\ref{line:ll:matching_tag}, or~\ref{line:ll:second_zero_tag} causes a tag to become strongly protected at $lin(c)$, and an \LL{} call that responds in \autoref{line:ll:final_return} does not strongly protect any tag, proving the first claim.
  If $c$ responds in \autoref{line:ll:final_return}, then $p$'s local variable $tag \neq 0$ and $tag' \notin \set{tag, 0}$.
  In particular, at point $t^\star \in (t, t')$, $p$ executes \autoref{line:ll:second_read}, in which it reads $tag'$ from $L[i].\Tag$.
  Thus, there is some point $t^\dagger \in (t, t^\star)$ at which $L[i].\Tag$ first changes to a value not in $\set{tag, 0}$.
  The only lines that can change $L[i].\Tag$ to a nonzero value are a \Swap{} in \autoref{line:swap:swap} or a successful \CAS{} in \autoref{line:sc2:first_cas} or~\ref{line:sc2:second_cas}.
  By \autoref{def:write:lin}, this operation is point $lin(c')$, for some call $c'$ which is either a \Swap{i, \cdot} call or a \SC{i, \cdot} call with a successful \CAS{} on $L[i]$.
  Moreover, since $c'$ and $c$ overlap, it follows that $c'$'s calling process $q \neq p$.
\end{proof}

\begin{theorem}[Linearizability]\label{thm:write:linearizable}
  The LL/SC implementation is linearizable according to the sequential specification, with $lin(c)$ as the linearization point for each call $c$.
\end{theorem}
This corresponds to Theorem 22 of~\cite{BHW2026a-full}, and the proof is mostly unchanged.
\begin{proof}
  Consider an execution of swappable LL/SC objects using this implementation.
  For each pending call $c$ in the execution such that $lin(c)$ is undefined according to \autoref{def:write:lin}, we remove $c$ from the linearization.
  For each \CL{} call invoked internally from an \LL{}, \SC{} or \VL{} call, we also remove it from the linearization.
  We construct the linearization from the remaining calls, sorting them by $lin(c)$.
  (Recall that \cleanup{} is a helper method, and so is not directly called in an execution.
  As such, it is not part of the linearization.)

  We prove that this linearization is correct, by induction over calls $c$ sorted by $lin(c)$.
  Specifically, we prove that each call $c$ returns a valid response and correctly affects the interpreted abstract state of the system.
  We inductively show that the interpreted value of LL/SC object $L_i$ is $L[i].\val$ before and after each linearization point.
  This is true initially since $L[i].\val$ is initialized to $u_i$, the application-dependent initial value for $L_i$.
  Recall from the sequential specification (\autoref{sec:write:seq_spec}) that the context bit $\ell_p$ of $L_i$ equals $1$ (i.e., $p$ has an open link to $L_i$) immediately before a given method call $c$ of a sequential execution if and only if
  \begin{itemize}
    \item an \LL{i} operation $c'$ is called by process $p$ before call $c$ in the sequential execution,
    \item no \Swap{i, \cdot} or successful \SC{i, \cdot} call (i.e., a call returning \True) by any process takes place in between $c'$ and $c$, and
    \item no \CL{i} call by $p$ takes place in between $c'$ and $c$.
  \end{itemize}
  It also follows from the sequential specification that the context bit only directly affects the return value of \SC{i, \cdot} and \VL{i} calls.
  As such, to prove correct behaviour of context bits, it suffices to show that \SC{i, \cdot} and \VL{i} calls return $\True$ if and only if the relevant context bit should be $1$ immediately before the respective call in the linearization.

  If $c$ is a \Read{i} or \LL{i} call, then its correctness follows from the fact that the call returns $L[i].\val$ at $lin(c)$.

  If $c$ is a \Swap{i, u} call, then its correctness follows from the fact that $L[i].\val$ changes from $v$ to $u$ at $lin(c)$, where $v$ is also the value that the call returns in \autoref{line:swap:return}.
  This is because $lin(c)$ takes place at the call's execution of \autoref{line:swap:swap}, which swaps $u$ into $L[i].\val$ and reads the previous value $v$, which is then returned.
  If an \SC{i, u} call $c$ returns \True, then $L[i].\val = u$ immediately after $lin(c)$, since by \autoref{def:write:lin} $c$ has a successful \CAS{} changing $L[i].\val$ to $u$ at point $lin(c)$.
  From the code and \autoref{def:write:lin}, $L[i].\val$ changes at point $t'$ only if there is a call $c'$ with $lin(c') = t'$, such that $c'$ is either a \Swap{i, u} call, or an \SC{i, u} call which returns \True.
  In both cases, $L[i].\val$ changes to $u$ at point $t'$.
  Therefore, if $c$ is an \SC{i, u} or \VL{i} call by process $p$, we only need to show that it returns \True if and only if the interpreted context bit $\ell_p = 1$, based on the three conditions listed earlier.

  Suppose that an \SC{i, u} or \VL{i} call $c$ by process $p$ does return \True, and we will show that $\ell_p = 1$ before $c$ according to the three conditions.
  By \autoref{lem:write:successful}(\ref{item:write:successful:strong_protection}), there is a point $t < lin(c)$ at which some tag $g$ becomes strongly protected for $L_i$ by $c$'s calling process $p$, and it remains strongly protected until at least $lin(c)$.
  By \autoref{def:write:protect}, only an \LL{i} call $c'$ by $p$ can cause $p$ to strongly protect $g$, and by \autoref{lem:write:ll_strong_protection}, $g$ becomes strongly protected at point $lin(c')$.

  Since a \CL{i} call by $p$ ends its strong protection of any tag for $L_i$, and $g$ remains strongly protected from $lin(c')$ until at least $lin(c)$, it follows that $p$ does not call \CL{i} in between $c'$ and $c$.
  By \autoref{lem:write:successful}(\ref{item:write:successful:no_others}) and~(\ref{item:write:successful:no_write}), no \SC{i, \cdot} call has a successful \CAS{} on $L[i]$ in the interval $(lin(c'), lin(c))$, and no execution of \autoref{line:swap:swap} takes place in that interval.
  Since an \SC{i, \cdot} call $s$ with a successful \CAS{} operation returns \True and has $lin(s)$ at the point of that \CAS{}, it follows that there is no such call $s$ with $lin(c') < lin(s) < lin(c)$.
  Similarly, since a \Swap{i, \cdot} call $w$ has $lin(w)$ at its execution of \autoref{line:swap:swap}, there is no such call $w$ with $lin(c') < lin(w) < lin(c)$.
  Therefore, the three conditions are satisfied and the interpreted context bit $\ell_p = 1$ immediately before $lin(c)$, as required.

  Now suppose that the interpreted context bit $\ell_p = 1$ immediately before call $lin(c)$ for an \SC{i, u} or \VL{i} call $c$ by process $p$.
  Then, by the three conditions, $p$ has an \LL{i} call $c'$ with $lin(c') < lin(c)$, such that $p$ does not directly call \CL{i} in between $c'$ and $c$.
  There is also no successful \CAS{} from an \SC{i, \cdot} call $s$ in \autoref{line:sc2:first_cas} or~\ref{line:sc2:second_cas} such that $lin(c') < lin(s) < lin(c)$.
  Similarly, there is no \Swap{} hardware instruction from a \Swap{i, \cdot} call $w$ in \autoref{line:swap:swap} such that $lin(c') < lin(w) < lin(c)$.

  Suppose without loss of generality that $c'$ is the last \LL{i} call by $p$ before $c$ (if the three conditions hold for an earlier \LL{i} call, then they also hold for $c'$).
  From \autoref{lem:write:ll_strong_protection}, some tag $g$ becomes strongly protected by $p$ at point $lin(c')$ (since $p$ responds from $c'$ before it invokes $c$).
  In order to use \autoref{lem:write:successful}, we need to show that this strong protection lasts until at least $lin(c)$, and in particular that $p$ does not indirectly call \CL{i} in between $lin(c')$ and $lin(c)$.
  (Recall that indirect calls of \CL{} are removed from the linearization).
  The possible indirect calls of \CL{i} are in lines~\ref{line:ll:reset_link},~\ref{line:ll:clear},~\ref{line:sc2:clear}, and~\ref{line:vl:clear}, so we will show that $p$ does not execute any of these lines in between $lin(c')$ and $lin(c)$ by deriving contradictions for each case.

  For \autoref{line:ll:reset_link}, since the line occurs before any possible line mapped to by $lin$ for \LL{}, it would have to be part of an \LL{i} call after $c'$, which contradicts that $c'$ is the last \LL{i} call by $p$ before $c$.
  For \autoref{line:ll:clear}, by the above arguments it should be part of call $c'$, but by \autoref{def:write:protect} and inspection of the code, no tag becomes strongly protected during $c'$, which contradicts that some tag does become strongly protected at point $lin(c')$.
  Line~\ref{line:sc2:clear} would be part of an \SC{i, \cdot} call, so we will show that $p$ does not make any such call in between $c'$ and $c$.
  If such a call were to return \True, this contradicts the assumption there is no \SC{i, \cdot} call which returns \True in the linearization in between calls $c'$ and $c$.
  If such a call were to return \False, then by the inductive hypothesis there is another \SC{i, \cdot} call $s$ which returns \True with $lin(c') < lin(s) < lin(c)$, and this also contradicts our assumption.
  This last argument also applies if $p$ executes \autoref{line:vl:clear} in between $lin(c')$ and $lin(c)$, since that line is only executed in a \VL{i} call which returns \False.

  We therefore have that a tag becomes strongly protected by $p$ for $L_i$ at point $lin(c')$ and remains so until at least point $lin(c)$.
  Since no successful \CAS{} from an \SC{i, \cdot} call and no \Swap{} hardware instruction from a \Swap{i, \cdot} call takes place in between $lin(c')$ and $lin(c)$, from \autoref{lem:write:successful}, $c$ does return \True.

  Since \CL{} only affects $\ell_p$, which itself only becomes visible via \SC{} and \VL{} calls which we have already covered, the proof is complete.
\end{proof}

Theorems~\ref{thm:write:history_independence},~\ref{thm:write:complexity}, and~\ref{thm:write:linearizable} together imply \autoref{thm:write:main}.

\section{Conclusion}\label{sec:conclusion}

We have shown how to add an unconditional and randomized wait-free \Swap{} operation to a QHI-preserving LL/SC implementation~\cite{BHW2026a}, using base objects that support swap, compare-and-swap, and fetch-and-increment/decrement, with the same asymptotic space and expected step complexity.
The changes require significant modifications to the technical analysis in order to prove that the results still hold with this larger set of supported operations.
Our result paves the way for designing more algorithms in the new field of history-independent concurrency.
A natural open problem is to implement more primitive operations in software that preserve history independence.

\bibliographystyle{plainurl}
\bibliography{pwliterature}

\end{document}